\documentstyle[11pt,aaspp4,psfig]{article}
\def\be{\begin{equation}}
\def\ee{\end{equation}}

\def\lambar{\lambda\llap {--}}

\def\lsim{\lower 2pt \hbox{$\, \buildrel {\scriptstyle <}\over
         {\scriptstyle \sim}\,$}}
\newcommand\gsim{\buildrel > \over \sim}
\begin{document}
\newcommand{\figureout}[3]{\psfig{figure=#1,width=7in,angle=#2} 
   \figcaption{#3} }

\title{Pulsar Polar Cap Heating and Surface Thermal X-Ray\\ 
 Emission I. Curvature Radiation Pair Fronts}

\author{Alice K. Harding\altaffilmark{1} and Alexander G. Muslimov\altaffilmark{2}}
\altaffiltext{1} {Laboratory of High Energy Astrophysics, NASA/Goddard Space  
 Flight Center, Greenbelt, MD 20771}
\altaffiltext{2} {Emergent Information Technologies, Inc., Space Sciences 
Sector, Upper Marlboro, MD 20774}  

\begin{abstract}
We investigate the effect of pulsar polar cap (PC) heating produced 
by positrons returning from the upper pair formation front.  Our calculations
are based on a self-consistent treatment of the pair dynamics and the effect
of electric field screening by the returning positrons.  We
calculate the resultant X-ray luminosities, and discuss the dependence 
of the PC heating efficiencies on pulsar parameters, such as 
characteristic spin-down age, spin period, and surface magnetic field 
strength. In this study we concentrate on the regime where the pairs 
are produced in a magnetic field by curvature photons emitted by 
accelerating electrons. Our theoretical results are not in conflict 
with the available observational X-ray data and suggest that the effect 
of PC heating should significantly contribute to the thermal X-ray 
fluxes from middle-aged and old pulsars. The implications for current and 
future X-ray observations of pulsars are briefly outlined.   
\end{abstract}


\section{INTRODUCTION}

X-ray emission has been detected from several dozen pulsars in observations
by ROSAT (Becker \& Trumper 1997), ASCA (Saito 1998) and Chandra 
(Zavlin et al. 2000).  In cases where good spectral measurements are available, 
the emission in the X-ray band
seems to have both thermal and non-thermal components.  While the non-thermal
components are most likely of magnetospheric origin, especially when they connect
smoothly to higher energy spectra, the origin of the thermal components is less 
clear because there are several mechanisms for production of thermal emission in
the soft X-ray band.  Pulsars younger than about $10^5$ yr are expected to have
significant neutron star (NS) cooling components and observed X-rays from these pulsars
may be consistent with standard cooling scenarios (Page 1998).  However, many older pulsars, 
including millisecond (ms) pulsars, beyond the age where cooling emission drops sharply,
have strong thermal X-ray emission. In this case, some heating mechanism(s) must be operating.
Two which have been proposed are heating by particles, accelerated in the magnetosphere,
flowing back to the NS PCs (PC heating: Ruderman \& Sutherland 1975, hereafter RS75; 
Arons 1981, hereafter A81) 
and internal heating by frictional forces in the crust (Shibazaki \& Lamb 1989).  

Because PC heating emission is intimately tied to the pulsar acceleration
mechanism, X-ray emission of older pulsars provides very strong constraints on
pulsar models.  Models which predict X-ray luminosities from PC heating which
are in excess of observed luminosities are not viable.  The earliest PC models,
based on vacuum gaps at the poles (RS75), predicted PC heating 
luminosities that were way above the first Einstein detections and upper limits on pulsars 
(Helfand et al. 1980).  However, predictions of the space-charge limited flow (SCLF) 
model of A81 were within all observed values.  Similarly, the PC heating predictions
of the first outer-gap models (Cheng, Ho \& Ruderman 1986), where particles are
accelerated in the outer magnetosphere, were well above the observed X-ray luminosities.
Later models allowed for a reduction of this huge PC X-ray emission by scattering
in a ``pair blanket" which forms near the surface (Wang et al. 1998).
It is clear that models with PC vacuum gaps and outer gaps, where half of 
all the particle acceleration energy returns to the surface to heat the PCs, 
will have severe problems preventing or hiding the excess X-ray emission.  
In this respect, SCLF models have a distinct advantage in that the number of 
particles that are accelerated downward to heat
the PC are only a small fraction of the number of primary particles that are
accelerated upward.  This is because the accelerating electric field (parallel to the 
magnetic field) arises not from a pure corotation charge, but from a slight 
imbalance between the corotation charge and the actual charge.
The field grows with height above the surface until electron-positron pairs from the
accelerated, upward moving electrons screen the field above a pair formation front (PFF).
A81, in his study of SCLF acceleration, found that the screening is primarily
due to positrons which turn around and accelerate backward to the stellar surface.  The 
number of returning (``trapped'') positrons required to screen the $E_{\parallel}$ is very small
compared to the number of primary particles (electrons).  In his studies, he assumed 
that pairs were produced only by curvature radiation (CR) photons.

We have begun an updated study of PC SCLF acceleration (Harding \& Muslimov 1998; 
hereafter HM98) based on the work of Muslimov \& Tsygan (1992; hereafter MT92), 
who found that including effects of general relativity produced a substantial 
difference in the parallel electric field induced above the surface of a NS.  
In particular, the effect of inertial frame dragging induces a field that 
is typically a factor of 50 times higher than that derived by A81, which is 
that induced by flaring of polar field lines alone.  This relativistic effect 
of inertial frame dragging and the consideration of radiation from inverse 
Compton scattering (ICS) of thermal X-rays from the NS surface by the primary 
electrons introduces profound changes in SCLF models.  We (HM98) found that 
pairs are produced by ICS photons from accelerating electrons at lower 
altitudes than are pairs from CR photons.  Pairs from ICS radiation could 
therefore screen the $E_{\parallel}$ before any CR photons are produced.  
We further found that positrons accelerating downward from ICS PFFs 
could potentially cause unstable acceleration if ICS pairs from positrons 
screened the $E_{\parallel}$ above the surface.  In our previous study, 
we computed self-consistently the height at which screening begins by assuming that 
the CR and ICS PFFs are located at the point where the first pair is produced.  
We also have given rough estimates for the returning positron flux and polar cap heating
luminosity expected from CR PFFs (Zhang \& Harding 2000).  However,
we did not model the structure of the PFF and thus could not determine
the flux of returning positrons as a fraction of the flux of primaries 
(the returning positron fraction) or even whether full screening occurs.  Many 
unanswered questions therefore remain, such as whether ICS photons can always 
screen the $E_{\parallel}$ and whether there are enough returning
positrons to produce pairs which screen $E_{\parallel}$ near the surface. 

To address these questions, as well as provide PC heating rates for the new
SCLF models, we have made a detailed and self-consistent study of the screening at
both CR and ICS PFFs.  We have found that while ICS photons are able to produce 
PFFs in nearly all pulsars, ICS radiation is able to completely screen the 
$E_{\parallel}$ only in pulsars with higher surface magnetic fields.  The
magnetic field value required for ICS screening is strongly dependent on the 
surface temperature of the NS.  For PC temperatures $T < 10^6$ K, pulsars with surface 
$B_0 \gsim 0.1\,B_{\rm cr}$, where $B_{\rm cr} \equiv 4.4 \times 10^{13}$ G is the critical
field strength, are capable of ICS screening, but for $T \gsim 10^6$ K, the required 
field decreases with temperature.  If the ICS photons are unable to 
completely screen the $E_{\parallel}$, then the primary electrons will keep accelerating
until they radiate CR photons which can produce a PFF.  In some cases, such as ms 
pulsars, incomplete ICS screening via non-resonant scattering can make a significant 
contribution to PC heating, especially for those pulsars where CR photons cannot 
produce any pairs. The results of these studies will be presented in two papers. 
This first paper will discuss CR screening and PC heating in lower field pulsars, 
i.e. those below the threshold for ICS screening.  In a second paper, we will discuss 
ICS screening in higher field pulsars, as well as partial ICS screening in older and ms 
pulsars, and will address the issues of lower PFFs and stability of ICS PFFs.  
In the present paper, we first discuss the SCLF electric field solution used and 
the conditions for the screening of this field.  We then present analytic estimates 
and numerical calculations of the returning positron fraction and PC heating luminosity 
for CR screening and compare our predictions with ROSAT observations. 

The most recent relevant study is that of Hibschman \& Arons (2001), who focus on
the determination of theoretical radio pulsar death lines.  Even though they
use the frame-dragging electric field in their treatment of primary electron 
acceleration and incorporate both CR and ICS photons in the pair production, our
approaches are essentially different.  For example, in their calculation of the 
PC heating they use a very rough estimate of the returning positron fraction (which 
is generally inaccurate and is the same as that given by MT92).  Also, they refer to 
the PFF as the location where the electric field is fully screened, whereas in this paper
we define the PFF as the location where pair production begins.  In the case of CR,
this distinction is minor since the screening scale is small compared to the PFF height, 
but in the case of ICS the distinction is very important since the screening scale 
is comparable to the PFF height.  Thus, the quantitative results of our studies are 
not directly comparable, although some of their conclusions are qualitatively
similar.  A more detailed comparison of our results with theirs will be given in
our next paper, where we will present our theoretical radio pulsar deathlines.

In Section 2, we describe the model we employ to calculate the 
accelerating electric field in the PC region of pulsars. We 
outline our main assumptions and approximations. The electrodynamic 
boundary conditions at the stellar surface and upper PFF are specified. 
In Section 3, we present our calculations of the returning positron fraction. 
In this paper our analysis is limited to the parameter space where the 
CR is the primary source of photons producing pairs in 
the magnetic field. We give our analytic estimates first. Then, we 
discuss our numerical treatment of electron-positron pair dynamics 
and effect of partial backflow of positrons. In Section 4, we calculate 
both analytically and numerically the thermal X-ray luminosities from 
pulsar PCs produced by precipitating positrons. Summary and conclusions 
are given in Section 5. 

\section{THE UNDERLYING ACCELERATION MODEL}
\subsection{Electric fields in the regime of space-charge limitation}

In this Section we shall summarize some of the general solutions 
pertaining to the electrodynamics of open field line regions of 
rotating NSs derived earlier (MT92, MH97, and HM98) and which will be 
exploited throughout this paper. We should note that these solutions 
are based on the assumption that a regime known as {\it space-charge 
limitation of current} occurs in the acceleration of electrons 
ejected from the PCs of a NS. The effect of {\it space-charge 
limitation} is well-known since the very first experimental and 
theoretical studies of the vacuum diodes, and detailed discussion of 
this effect can be found in any textbook on electronic devices. 
In Section 2.2 we will briefly outline some of the features 
of the {\it space-charge limitation} specific to the physical 
conditions of the pulsar PC regions. 

The general relativistic Poisson's equation describing the electric 
potential distribution (in the corotating frame of reference) within 
the region of open field lines in the magnetosphere of a rotating 
NS reads (MT92)
\be
\nabla \cdot \left( {1\over \alpha } \nabla \Phi \right) 
= - 4 \pi ( \rho - \rho _{\sc {GJ}} ) ,
\label{Poisson'sEqn}
\ee
where $\rho $ is the space charge density, $\rho _{\sc {GJ}}$ is the 
general relativistic analog of the Goldreich-Julian (GJ) space charge density, 
$\alpha = \sqrt{1-\epsilon /\eta }$ is the redshift function, 
$\epsilon = r_g/R$, $r_g$ is the gravitational radius of the NS, $R$ is 
the stellar radius, $\eta = r/R$ is the dimensionless radial coordinate, 
$\kappa = \epsilon I/MR^2$, $I$ and $M$ are the moment of inertia and mass 
of the NS, respectively. The differential operators such as gradient, 
divergence, etc. should be taken in corresponding curvilinear coordinates. 
Eqn. (\ref{Poisson'sEqn}) takes into account both the effect of 
{\it dragging of inertial frames of reference } and 
{\it gravitational redshift}. The former dramatically affects the 
electrodynamics of a NS, while the latter rather moderately modifies 
the corresponding electrodynamic quantities. 

The study of particle acceleration within the region of open magnetic 
field lines in pulsars is grossly facilitated by the fact that for all 
known pulsars (including ms pulsars) the angular size of the 
PC is less than $\sim 0.2/\sqrt{P/3~{\rm ms}}$ radian 
($P$ is a pulsar spin period). Thus, we can use a so-called 
small-angle approximation, which proves to be very satisfactory for 
most relevant problems. In this approximation, for general relativistic 
dipolar magnetic field,
\be
\rho _{\sc {GJ}}(\eta , \xi , \phi ) = - \sigma (\eta ) 
\left[ \left( 1 - {{\kappa }\over {\eta ^3}} 
\right) \cos \chi + {3\over 2} 
\theta (\eta ) H(\eta )\xi \sin \chi \cos \phi \right] ,
\label{rhoGJ}
\ee
\be   
\sigma (\eta ) = {{\Omega B_0}\over {2 \pi c \alpha \eta ^3}} 
{{f(\eta )}\over {f(1)}},
\label{sigma}
\ee
where $\Omega $ is the angular velocity of the NS rotation, $B_0$ is the 
surface value of the magnetic field strength at the magnetic pole and 
$\chi$ is the pulsar inclination angle. 
Here we use the dimensionless radial coordinate $\eta $ ($\equiv r/R$) and 
characteristics describing field-streamlines $\xi $ 
$ = \theta /\theta (\eta )$ [which is a magnetic colatitude scaled by 
the half-opening angle of the polar magnetic flux tube, 
$\theta (\eta ) = \theta _0 \sqrt{\eta f(1)/f(\eta )}$, 
where $\theta _0 = \sqrt{\Omega R/cf(1)}$ is PC half-angle], 
$\phi $ is magnetic azimuthal angle, and the functions $f(\eta )$ and 
$H(\eta )$ are factors accounting for the static part of the curved 
space-time metric [see Eqns (A11), (A12) in Appendix].  

In this paper we continue our study of a regime of {\it space-charge 
limitation of current}, which allows us to derive an explicit 
expression for a steady-state distribution of space charge density 
(see MH97 for details) 
\be
\rho (\eta , \xi , \phi ) = - \sigma (\eta ) \left[ (1 - \kappa ) 
\cos \chi + {3\over 2} \theta _0 H(1) \xi \sin \chi \cos \phi \right] .
\label{rho}
\ee
Using expressions (\ref{rhoGJ})-(\ref{rho}) and appropriate boundary 
conditions we can solve Eqn. (\ref{Poisson'sEqn}). In this paper we 
will be mostly interested in the  solutions for the accelerating 
electrostatic potential (field) valid for altitudes less than one stellar 
radius from the PC surface i.e. for $z = \eta -1 < 1$, even though 
their generalization for $z > 1$ proves trivial (HM98). Thus, for 
$z < 1$ the general solution for the potential can be expressed as 
\be 
\Phi (z, \xi , \phi ) = \Phi _0 \theta _0^3 \sqrt{1-\epsilon }  
\sum _{i=1}^{\infty } \left[ A_i J_0(k_i \xi ) \cos \chi + 
B_i J_1({\tilde k}_i \xi ) \sin \chi \cos \phi \right],
\label{Phi_gen}
\ee
where
\be
A_i(z) = {3\over 2} \kappa {\cal F}_i(z,\gamma _i) 
\left[ {8\over {k_i^4J_1(k_i)}} \right],
\label{A_i}
\ee
\be
B_i(z) = {3\over 8} \theta _0 H(1) \delta (1){\cal F}_i(z,{\tilde 
{\gamma }} _i)\left[ {16\over {{\tilde k}_i^4J_2({\tilde k}_i)}} \right],
\label{B_i}
\ee
and
\be
\gamma _i = { {k_i} \over {\theta _0 \sqrt{1-\epsilon } }}.
\label{gamma_i}
\ee
Here $\Phi _0 \equiv (\Omega R/c)B_0 R$ is the dimensional factor 
which is nothing but an order-of-magnitude maximum value of electrostatic 
potential generated by a magnetized globe (with dipolar external magnetic 
field) rotating in vacuo (see Deutsch 1955); $k_i$ and ${\tilde {k}}_i$ are 
the positive zeros of the Bessel functions $J_0$ and $J_1$, respectively, 
indexed in ascending order; $\delta(\eta )$ is another correction factor 
[see Eqn. (A13)] accounting for the gravitational redshift effect, 
with $H(1) \delta (1) \approx 1$ (see e.g. HM98); and $\theta _0$ is 
defined right after Eqn. (\ref{sigma}). The explicit expressions for the 
function ${\cal F}_i$ are determined by the solution of Eqn. (\ref{Poisson'sEqn}) 
and appropriate electrodynamic boundary conditions. The solutions 
for the function ${\cal F}_i$ relevant to our study will be presented 
in \S ~2.3 and the Appendix. In this paper we do not need and therefore 
do not present the explicit general-relativistic expressions for the 
magnetic field, but we shall refer an interested reader to our previous 
publications (MH97, HM98). Finally, our results will be based on the 
electrodynamic solutions of Eqn. (\ref{Poisson'sEqn}) subject to 
the standard Dirichlet boundary conditions.  

\subsection{Modification of space-charge limitation by returning positrons}

In this Section we shall discuss the space-charge limitation of current 
taking into account the returning positron flux from the upper PFF. 
The idea of space-charge limitation of flow was in fact introduced 
into pulsar physics by Sturrock (1971): "{\it What happens to the 
positrons produced by the electron-positron avalanche? If they were all 
returned to the surface, as one might expect by consideration of the 
electric field, the resulting space charge would reverse the sign of 
the electric field at the surface, cutting off the flux of primary 
electrons"}. Later on this idea was investigated by Michel (1974, 1975), 
Tademaru (1973), and Cheng and Ruderman (1977), and then substantially 
quantified in the calculation of the electric fields produced by the 
relativistic electron beam within the region of open field lines in 
a NS magnetosphere (Arons \& Scharlemann, 1978; Arons 1983, hereafter A83). 
They found that only a small fraction of positrons (e.g. $\sim \theta _0^2$ 
of the negative charge density already present in the electron beam, 
as estimated by Arons \& Scharlemann) are returned relative to the 
primary electron beam. Note that the calculation of the fraction of 
positrons returning from the upper PFF is far from being a trivial 
problem. In this paper we attempt to address this problem again, 
taking into account most recent updates of the physics involved and 
revising some of the underlying assumptions.

As a result of injection of backflowing positrons into the initial 
electron beam, an excessive negative space charge unavoidably builds up 
which lowers the electrostatic potential above the PC surface. The 
lowering of potential above the surface and forcing its gradient to 
reverse sign at the very surface would inhibit electron emission, 
thus resetting the electric field to zero at the surface. The 
fundamental difference between the laboratory vacuum diode and PC 
voltage generator is that in the former the thermionic cathode is the 
only emitter of electrons, whereas in the latter the PC surface, 
serving as a cathode, both emits electrons and collects positrons from 
the upper PFF. Thus, for a vacuum diode the space-charge limitation of 
current is solely determined by the cathode, while for a NS PC this 
regime of operation depends on the upper PFF as well. Because of the 
intrinsic  limitation of the total current by the GJ value, the injection 
of counterstreaming positrons suppresses the primary electron emission: 
the primary electrons should be ejected from the stellar surface with a 
density somewhat less than the GJ charge density to keep $E_{\parallel }=0$ 
at the surface. This effect provides a negative feedback between the 
primary electron ejection and flux of returning positrons, since the 
primary electrons ultimately determine (through the electrostatic potential 
drop they generate above the PC surface) the fraction of returning positrons. 

Now, using very basic arguments, we would like to demonstrate how the 
accelerating electric field is related with the fraction of returning 
positrons. In our previous paper (see HM98) we quantitatively explored 
the effect of rescaling of the lower boundary (e.g. stellar surface) 
caused by the formation of a lower pair front in the regime of space-charge 
limitation. The returning positrons add to the space charge density of 
primary electrons and lift up  the boundary at which the $\rho = \rho _{\sc GJ}$ 
condition is satisfied, similar to the effect produced by the formation 
of lower pair front. Let us use this reasoning and express the space charge 
density of a primary beam in terms of effective (rescaled) stellar radius, 
$R_E$. For simplicity, we illustrate this by employing solution (\ref{rho}) 
at $\chi \approx 0$. 
The expression for the space charge density at the rescaled stellar surface 
then reduces to 
\be
\rho = - \sigma (1- \eta _{\ast }^3 \kappa) ,
\ee
where $\eta _{\ast } = R/R_E$ ($\leq 1$). \\
If there were no returning positrons ($\eta _{\ast }\equiv 1$) we would 
simply have 
\be
\rho _{-} = - \sigma (1-\kappa) .
\ee
The returning positrons perturb the primary electron beam causing 
extraction of less electrons than in the case of unperturbed electron 
beam and effectively mimic electrons missing in a beam. Assuming that 
the regime of space-charge limitation is maintained, the following 
condition should be satisfied at the surface               
\be 
\rho = \rho _{-} + \delta \rho _{+}, 
\ee
where $\delta \rho _{+}$ is a small perturbation of (and of the same 
sign as) the electron space charge density caused by returning positrons.
Introducing the dimensionless fraction of returning positrons, $x_{+}$, 
so that $\delta \rho _{+} = - \sigma \kappa x_{+}$, from 
the above equation we get that $\eta _{\ast } = (1-x_{+})^{1/3}$. 
Thus the positrons returning from the PFF reduce the accelerating 
electrostatic potential (field), so that the latter gets factorized by 
(see also HM98) 
\be    
\eta _{\ast }^3 = 1 - x_{+}.
\ee
It is now clear that the net effect of the injection (e.g. from the upper 
PFF) of counterstreaming positrons into primary electron beam is a 
weakening of the source of effective charge, or, equivalently, the 
transforming of general-relativistic parameter $\kappa $ into  
$(1-x_{+})\kappa $. Note that $x_+ \ll 1$ in most cases. Thus the final 
result does not depend on the particular value of rescaled stellar radius 
and illustrates the dynamics of the feedback between accelerating electric 
field and fraction of returning positrons. Namely, the returning positrons 
tend to reduce the accelerating electric field and therefore suppress the 
ejection of primary electrons. The starving electron ejection is compensated 
by returning positrons, which mimic the outflowing electrons, thus recovering 
the original potential drop. Obviously, the increase in the fraction of 
backflowing positrons reduces the efficiency of pair production by primary 
electrons which may eventually cut off the supply of positrons themselves. 
One can expect some demand-supply balance to be established in the 
beam between the densities of primary electrons and backflowing positrons. 
Thus, we are now in a position to calculate acceleration of primary electrons 
and returning positrons produced at the PFF in a self-consistent way, taking 
into account the weakening of the primary beam by returning positrons. It is 
important that, after the appropriate parameters that favour the steady-state 
regime are found, we can use essentially the same expressions for the 
accelerating electric field/potential as in the case of a pure electron beam, 
simply because in a steady-state electrodynamic solution the returning 
positrons are undistinguishable from outflowing electrons. Then, the problem 
reduces to finding (through numerical iterations) a stable solution 
(see \S~3.2.2 for details).

\subsection{Screening condition at the upper PFF and accelerating electric field}

In this paper we derive an appropriate solution for the accelerating electric field, 
and use this solution for a self-consistent analysis of a steady-state flow 
of charged particles (electrons and positrons) from the PC surface up through the 
PFF. We assume that both the PC surface and surface formed by the last open magnetic 
field lines are equipotential ($\Phi = 0$). We must note that, within the context 
of the lower boundary condition, the PC surface is actually either the upper boundary 
of the stellar atmosphere or the bare stellar crust. Since the free emission of 
charges from the PC seems to be more favourable than charge starvation above the PC, 
the electron current should be consistent with the global magnetospheric current. 
We assume that in a steady-state regime the rate of ejection of electrons from the 
PC surface is approximately equal to the rate of inflow of electrons from beneath 
the surface. It is this basic condition that makes the ejection of 
electrons occur without producing significant polarization of charges at the 
PC surface. Note that, in a static model ignoring the global current closure 
the free emission of electrons from the PC would rather result in a filling of 
the polar magnetic flux tube with electron-ion plasma and any electron acceleration 
will be halted. 

The upper boundary condition also needs to be discussed in detail, simply because 
it proves to be crucial for the calculation of the fraction of returning positrons 
from the upper PFFs. This condition was extensively discussed in the literature 
(see e.g. Arons \& Scharlemann, 1979; A83) and it was suggested to set 
$E_{\parallel } = 0$ and $\nabla _{\parallel } \cdot E_{\parallel } = 0 $ at the 
upper PFF. The problem is that these two conditions are formally inconsistent, 
as we shall demonstrate in the next paragraph. This inconsistency should not be 
ignored in any more or less realistic model simply because these two conditions 
cannot be satisfied not only at the same point but even at two points separated 
by a distance of order of the PFF thickness. 

The formal condition assumed previously that $\rho = \rho _{\sc {GJ}}$ at the 
upper PFF (at $\eta = \eta _0$) means that in the very vicinity of the PFF the 
radial part of electrostatic potential should satisfy the following equation 
[see e.g. MH97, Eqn. (48)] 
\be
{{\partial ^2 P}\over {\partial \eta ^2}} + 2 {{\partial P}
\over {\partial \eta }} - a^2 P = 0 ,
\label{P}
\ee
where $2/a = \theta (\eta _0) \eta _0 \sqrt{1-\epsilon /\eta _0} 
\sim r_{pc}/R$ ($r_{pc}$ is the PC radius).
The non-trivial solution of this equation reads 
\be
P = C \exp (- \eta ) \sinh [ \sqrt{1+a^2}(\eta _0 - \eta )], 
\ee
where $C = const$.
The boundary conditions $E_{\parallel} = 0$ and $\nabla _{\parallel }
\cdot E_{\parallel} = 0$ at $\eta = \eta _0$ are simply reduced 
to the following conditions for the function $P$ 
\be
{{\partial P}\over {\partial \eta }} = 0~~~~~{\rm and}~~~~~
{{\partial ^2 P}\over {\partial \eta ^2}} + 2 {{\partial P}
\over {\partial \eta }} = 0,
\ee
respectively. One can easily verify that the first of the above conditions 
is not satisfied. This means that it is incorrect to assume that both 
$\nabla _{\parallel } \cdot E_{\parallel} = 0$ and $E_{\parallel } = 0$ are 
satisfied at the same radial coordinate. The reason is that condition 
$\nabla _{\parallel} \cdot E_{\parallel} = 0$ actually implies by 
Eqn. (\ref{P}) that the electrostatic potential also vanishes at the 
upper PFF. However, the latter by no means requires that $E_{\parallel}$ 
vanishes at the same boundary. Moreover, as it is illustrated in Appendix 
[see Eqns (A2), (A3), and (A7)-(A10)], the vanishing of both 
$\nabla _{\parallel }\cdot E_{\parallel}$ and $E_{\parallel }$ is very 
unlikely even within entire PFF. In general, specifying both Dirichlet 
and Newmann conditions (e.g. $\nabla _{\parallel } \cdot E_{\parallel} = 0$ 
and $E_{\parallel } = 0$) overdetermines the problem and leads to 
there being no solution. 

In our solution (which is essential for the kind of problem we discuss 
in this paper) we will only require that beyond the upper PFF, 
\be
\Phi \rightarrow \Phi _{\infty} = const \neq 0,~~~~~{\rm as}~~~~~
\rho \rightarrow \rho _{\sc {GJ}} ,
\label{Phi_pff}
\ee
where $\Phi _{\infty }$ is the potential at altitudes well above the 
upper PFF. This condition implies that there is no perfect adjustment 
of the effective space charge density to the GJ space charge density 
at the very onset of the upper PFF. When $E_{\parallel }$ has dropped 
to a low enough value that positrons are no longer able to turn around, 
then there is still non-zero $\Phi $ but no additional returning flux. 
If, at the PFF, the GJ space charge were exactly compensated by the 
effective space charge of primary electrons and electron-positron plasma, 
then the electrostatic potential would drop to zero at the PFF. The 
vanishing of potential at the PFF suppresses the accelerating electric 
field and even causes it to reverse its sign well below the PFF. The 
grounding of potential at the PFF would therefore dramatically affect 
the acceleration of primary electrons: the electron beam will be cut off 
from the PFF by a layer with a reverse polarity of the electric field, 
and the primary electrons will no longer be able to maintain the PFF 
they would produce if there was no strong negative feedback under 
discussion. As a result, it is very unlikely that a stable upper PFF 
would be established at all, if the condition $\rho = \rho _{\sc {GJ}}$ 
sets in at the PFF. We suggest that requirement (\ref{Phi_pff}) is 
more favorable for a non-disruptive regime of acceleration of primary 
electrons and formation of upper pair fronts. This requirement allows 
the parallel component of the electric field to partially penetrate 
into the relativistically moving pair plasma, with the bulk of the 
electron-positron pairs streaming out as a quasi-neutral beam. 
We therefore relax the intuitive requirement that relativistically 
moving electron-positron pairs completely screen out the electric field 
at or right above the PFF. A self-consistent treatment of the dynamics 
of primary electrons and pairs in a nearly screened electric field, as  
will be discussed  below, may provide us with a reasonable estimate 
of the fraction of returning positrons. 

Using general expression (\ref{Phi_gen}) we can derive the formulae for the 
accelerating potential (field) for $0 < z < 1$. This solution has the property 
that $E_{\parallel }$ saturates  ($E_{\parallel } \rightarrow const$) 
for $r_{pc}/R << z < 1$, before it declines approximately as $\eta ^{-4}$ 
(see MT92, MH97, HM98). In this case (no explicit upper boundary) the 
solution for the function ${\cal F}_i(z,\gamma _i)$ is 
\be
{\cal F}_i = \gamma _i z + \exp(-\gamma _iz) - 1,
\label{F}
\ee
and the corresponding formulae for the potential and field read
\begin{eqnarray}
\Phi (z, \xi , \phi ) & = & {3\over 2} \Phi _0 {{\Omega R }\over c} 
{1\over f(1)} \left\{ \kappa \left[ 8\sum _{i = 1}^{\infty} 
{{J_0(k_i\xi )}\over {k_i^3 J_1(k_i)}}
\left((\exp(-\gamma _i z) -1)/\gamma _i + z\right) \right] 
\cos \chi + \right. \nonumber\\
&  &  \left. {1\over 4} \theta _0 H(1)\delta(1) \left[ 16\sum _{i = 1}
^{\infty}{{J_1({\tilde {k}}_i\xi )}
\over {{\tilde {k}}_i^3 J_2({\tilde {k}}_i)}}
\left( (\exp(-{\tilde{\gamma }_i}z) -1)/{\tilde {\gamma }}_i + z\right) 
\right] \sin \chi \cos \phi \right\}, 
\label{Phi_acc}
\end{eqnarray}
\begin{eqnarray}
E_{\parallel }(z,\xi ,\phi ) & = & - {3\over 2} E_0 {{\Omega R }\over c} 
{1\over f(1)} \left\{ \kappa \left[ 8\sum _{i = 1}^{\infty} 
{{J_0(k_i\xi )}\over {k_i^3 J_1(k_i)}}
\left(1 -\exp(-\gamma _i z)\right) \right] \cos \chi + \right. \nonumber\\
&  &  \left. {1\over 4} \theta _0 H(1)\delta(1) \left[ 16\sum _{i = 1}
^{\infty}{{J_1({\tilde {k}}_i\xi )}
\over {{\tilde {k}}_i^3 J_2({\tilde {k}}_i)}}
\left(1 -\exp(-{\tilde{\gamma }_i}z)\right) \right] \sin \chi 
\cos \phi \right\}, 
\label{E_acc}
\end{eqnarray}
where $E_0 \equiv \Omega R B_0/c = \Phi _0 / R$. 

Simple expressions can be derived from the above formulae in 
some limiting cases. For example, for $\gamma z \ll 1$ ($z \ll r_{pc}/R$) 
we get 
\begin{eqnarray}
\Phi (z, \xi , \phi ) & = & 1.6 \Phi _0 
\left( {{\Omega R}\over c} \right) ^{1/2} 
{z^2 \over {\sqrt{(1-\epsilon )f(1)}}} 
\left[ \kappa \left( 1 - \xi ^{2.19} \right) ^{0.705} 
\cos \chi + \right. \nonumber\\
& & \left. {3\over 8} H(1) \delta (1) \theta _0 \xi ^{1.015} 
\left( 1 - \xi ^2\right) ^{0.65} \sin \chi \cos \phi \right] ,
\label{Phi_unsat}
\end{eqnarray}
\begin{eqnarray}
E_{\parallel }(z, \xi , \phi ) & = & - 3.2 E_0 
\left( {{\Omega R}\over c} \right) ^{1/2} 
{z \over {\sqrt{(1-\epsilon )f(1)}}} 
\left[ \kappa \left( 1 - \xi ^{2.19} \right) ^{0.705} 
\cos \chi + \right. \nonumber\\
& & \left. {3\over 8} H(1) \delta (1) \theta _0 \xi ^{1.015} 
\left( 1 - \xi ^2\right) ^{0.65} \sin \chi \cos \phi \right] .
\label{E_unsat}
\end{eqnarray}

In derivation of the above expressions we have used the formulae 
\be
\sum _{i = 1}^{\infty }{{J_0(k_i x)}
\over {k_i^2 J_1(k_i) }} 
\approx {4\over 15}\left( 1 - x ^{2.19} \right) ^{0.705} ,
\ee
\be
\sum _{i = 1}^{\infty }{{J_1({\tilde {k}}_i x)}
\over {{\tilde {k}}_i^2 J_2({\tilde {k}_i}) }} 
\approx {1\over 5} x^{1.015} \left( 1 - x^2 \right) ^{0.65} ,
\ee
fitting the results of numerical summation with accuracy better than 
$2\% $. 
For $z \gg r_{pc}/R$ ($\gamma z \gg 1$) we get 
\be
\Phi (z,\xi , \phi ) = {3\over 2} \Phi _0 {{\Omega R }\over c} 
{z \over {f(1)}}(1 - \xi ^2)\left[ \kappa \cos \chi + 
{1\over 4} \theta _0 H(1)\delta(1) \xi \sin \chi \cos \phi \right], 
\label{Phi_sat}
\ee
\be
E_{\parallel }(z,\xi , \phi ) = - {3\over 2} E_0 {{\Omega R }\over c} 
{1\over {f(1)}} (1 - \xi ^2)\left[ \kappa \cos \chi + 
{1\over 4} \theta _0 H(1)\delta(1) \xi \sin \chi \cos \phi \right]. 
\label{E_sat}
\ee
Before we proceed with our modeling of field screening and calculation of 
the fraction of returning positrons, let us introduce a dimensionless 
minimum height $z_0$ ($\equiv \eta _0 - 1$) at which a first pair is 
produced . Then for our modeling we will need the expression for the 
electric field applicable to the region of pair formation. Since the 
penetration depth of the electric field into the pair plasma (or the 
characteristic length scale of the electric field screening) is much 
less than the characteristic length scale of the accelerating electric 
field [e.g. Eqn. (\ref{E_sat})], the electric field in the pair region 
can be satisfactorily described by the formula 
\be
E_{\parallel }^{sc}(z \gsim z_0) = E_{\parallel }^{acc}(z_0) 
\exp [-(z-z_0)/\Delta _s] ,
\label{E_pff} 
\ee
where $E_{\parallel }^{acc}(z_0)$ is the accelerating electric 
field given by (\ref{E_acc}) and calculated at $z = z_0$, $\Delta _s$ 
is the characteristic length scale of field screening.

\section{CALCULATION OF RETURNING POSITRON FRACTION}
\subsection{Analytic estimate}

Let us estimate very roughly the fractional density of positrons flowing 
back from the PFF. We shall restrict ourselves to the case where the 
pair creation is mostly determined by the CR photons. 
The condition that the positrons with energy $\varepsilon _+$ (in 
$mc^2$) turn around within the PFF can be written as 
\be
e |E_{\parallel }| \Delta _s = \varepsilon _+ mc^2,
\label{U-turn1}
\ee
where $E_{\parallel }$ is the accelerating electric field evaluated 
at the PFF, $\Delta _s$ is the same as defined in Eqn. (\ref{E_pff}), 
and $e = |e|$ is the elementary electron charge.  
The characteristic energy of turning around positrons can be estimated 
as 
\be
\varepsilon _+ \approx {1\over 2} \varepsilon _{\gamma } 
{n_e \over n_+} {{dN_{\gamma }}\over {dN_e}},
\label{eps_+}
\ee
where $\varepsilon _{\gamma }$ is the characteristic energy of 
pair-producing curvature photons, $n_e$ is the number density 
of primary electrons, $n_+$ is the number density of turning around 
positrons, and $dN_{\gamma }/dN_e$ is the number of photons (per 
primary electron) producing those pairs whose positrons turn around 
and flow back to the surface. 
Note that 
\be
\varepsilon _{\gamma } {{dN_{\gamma }}\over {dN_e}} = 
p(\varepsilon _{\gamma }) {{\Delta \varepsilon _{\gamma }}\over hc} 
\Delta _s,
\label{U-turn2}
\ee
where $p(\varepsilon ) = (\sqrt {3}e^2/2\pi \rho _c )\gamma 
F(\varepsilon /\varepsilon _{cr})$ is the spectral power of CR 
(see e.g. Landau \& Lifshitz, 1975), $\Delta \varepsilon _{\gamma }$ 
is the characteristic interval of energies in the CR spectrum from 
which the pair-producing photons generate positrons with the 
characteristic energy $\sim \varepsilon _+$. Using Eqns 
(\ref{U-turn1})-(\ref{U-turn2}) we can write
\be
{{n_+}\over {n_e}} \approx {1\over 2} p(\varepsilon _{\gamma }) 
{{\Delta \varepsilon _{\gamma }}\over h} {mc \over {e|E_{\parallel }|}},
\ee
which translates into
\be
{{n_+}\over {n_e}} \approx 5 \times 10^{-2} \Lambda F(\varepsilon 
_{\gamma } /\varepsilon _{cr}) 
\left( {{2e\gamma ^4}\over {3\rho _c^2 |E_{\parallel }|}} \right),
\label{n_+/n_e}
\ee
where $\varepsilon _{cr} = 3 \lambar _c \gamma ^3 / 2 \rho _c $ 
($\lambar _c \equiv \hbar /mc = 3.9 \times 10^{-11}$ cm is the 
Compton wavelength) is the critical energy of the curvature spectrum. 
The factor $\Lambda $ ($\equiv \Delta \varepsilon _{\gamma } /\varepsilon _{cr}$) 
in Eqn. (\ref{n_+/n_e}) takes into account the fact that only photons with 
approximately Gaussian distribution around some characteristic energy 
effectively produce pairs, with $\Delta \varepsilon _{\gamma }$ being the 
spectral interval of photons that produce returning positrons. The factor 
$F(x)$ [$= x \int _x^{\infty } K_{5/3}(z)dz \approx \sqrt{ \pi x/2} \exp (-x)$, 
if $x >>1$, where $K_{5/3}$ is the modified Bessel function of order 5/3] accounts 
of the fact that returning positrons are produced by CR photons with energies greater 
than $\varepsilon _{cr}$. Finally, factor 
$2e\gamma ^4/3\rho_c^2|E_{\parallel }| \leq 1 $ 
describes the efficiency of emission of CR by accelerating 
electron. It reaches a maximum in the so-called "saturation" regime where 
the power of electrostatic acceleration is almost exactly compensated by 
the CR losses. If we adopt in Eqn. (\ref{n_+/n_e}) $\Lambda \approx 3$, 
and $\varepsilon _{\gamma } \sim (3-6) \varepsilon _{cr}$ (this range 
for $\varepsilon _{\gamma }$ of photons producing returning positrons agrees 
with that resulting from our numerical simulations, with the characteristic 
value of $\varepsilon _{\gamma }$ being a factor of 2-3 smaller than the 
mean energy of pair-producing photons given by Eqn. [29] in HM98), we get 
\be
{{n_+}\over {n_e}} \approx (0.001-0.02)\left( {{2e\gamma ^4}
\over {3\rho _c^2 |E_{\parallel }|}} \right).
\label{n_+/n_e-num}
\ee
It is interesting, that the fractional density of returning positrons 
given by the above rough formula is practically independent of pulsar 
parameters, except for the dimensionless factor in brackets. The latter 
also tends to be close to its maximum value of unity. The reason is 
that in most physically interesting situations $\gamma $ should be 
well tuned to allow magnetic pair production, which occurs at or near 
the saturation part of the acceleration curve, where the further 
acceleration of electrons would be suppressed by the CR losses, 
i.e. where the condition 
$e |E_{\parallel }| \sim 2 e^2 \gamma ^4 /3 \rho _c^2$ is roughly 
satisfied. 
  
Even though formula (\ref{n_+/n_e-num}) gives the right range for the 
fractional density of returning positrons, it is based on a simplified 
mapping between the spectrum of pair-producing CR photons and the 
distribution function of returning positrons. Its main deficiency is 
that it requires {\it a~priori} knowledge (say, based on numerical 
simulation) of parameters $\Lambda $ and $\varepsilon _{\gamma }$. 
Also, it does not reflect the fact that at the upper PFF the maximum 
density of returning positrons is limited by the value of 
$|\rho _{\sc GJ} - \rho |$ and is therefore dependent 
on the altitude of the PFF. In a steady state one can estimate the 
maximum value of the density of returning positrons as one half of 
the difference $|\rho _{\sc GJ}(z_0) - \rho (z_0)|$. Thus, using 
expressions (\ref{rhoGJ}) and (\ref{rho}) for $\rho _{\sc GJ}$ and 
$\rho $ we can come up with an alternative formula  
\be
{{\rho _+(z_0)}\over {\rho _{\sc GJ}(z_0)}} = {1\over 2} 
\left[ 1 - {{\rho (z_0)}\over {\rho _{\sc GJ}(z_0)}} \right] 
\approx {3\over 2} {{\kappa }\over {1-\kappa}} z_0 .
\label{rho+/rhoGJ}
\ee
A similar formula has been used by Zhang \& Harding (2000) in their 
study of the X-ray luminosities from the spinning-down pulsars. 

Let us perform numerical estimates of $z_0$ based on formulae 
(\ref{E_unsat}) and (\ref{E_sat}) for the accelerating electric field. 
To make the resulting estimates as compact as possible, we shall adopt 
the following parameters: $\xi = 0.5$,  $\cos \chi = 1$, and 
$\kappa = 0.15$. This is still a justified simplification, since the 
resultant expressions will have rather weak dependence on these 
parameters. Then, Eqns (\ref{E_unsat}) and (\ref{E_sat}) reduce to 
\be
E_{\parallel 6} = 42 {{B_{12}}\over {P_{0.1}^{3/2}}} z ,
\label{EI}
\ee
and 
\be
E_{\parallel 6} = 0.5 {{B_{12}}\over {P_{0.1}^2}} ,
\label{EII}
\ee
respectively. Here $E_{\parallel 6} \equiv E_{\parallel }/10^6$ esu, 
$B_{12} = B_0/10^{12}$ G, and $P_{0.1} = P/0.1~s$. 

The altitude of the PFF above the stellar surface can be estimated as 
[see e.g. HM98, Eqn. (1)]
\be 
S_0 = min [ S_a(\gamma _{min}) + S_p(\varepsilon _{min})],
\label{S0_1}
\ee
where $S_a(\gamma _{min})$ is the distance required to accelerate the 
particle until it can emit a photon of energy $\varepsilon _{min}$ and 
$S_p(\varepsilon _{min})$ is the photon pair-attenuation length. 
Thus, using expressions (\ref{EI}), (\ref{EII}) we arrive at 
\be
S_0 = min (S_a + B_{\gamma }/\gamma ^3 ),
\label{S0_2}
\ee
where
\be
S_a^{\sc {I,II}} = \left\{ \begin{array}{ll}
    A_{\gamma }^{\sc I} \gamma ^{1/2}  & {\rm if}\: E_{\parallel }~{\rm is 
        ~given~by}~(\ref{E_unsat}), \\
    A_{\gamma }^{\sc II} \gamma & {\rm if}~E_{\parallel}~{\rm is~given~by}~ (\ref{E_sat}).
\end{array} 
\right.
\ee
Here $A_{\gamma }^{\sc I} = 9.1\cdot P_{0.1}^{3/4}/B_{12}^{1/2}$ , 
$A_{\gamma }^{\sc II} = 3.3\cdot 10^{-3} P_{0.1}^2/B_{12}$, 
and $B_{\gamma } = 1.95\cdot 10^{26} P_{0.1}/B_{12}$. Note that this  
expression for $B_{\gamma }$ is valid for $B_{12} \leq 4.4$. 
Equation (\ref{S0_2}) has a minimum at the value of Lorentz factor 
\be
\gamma_{min} = 10^7 \left\{ \begin{array}{ll}
    2.9~P_{0.1}^{1/14}B_{12}^{-1/7} & {\rm if}\: E_{\parallel } ~
		{\rm is~given~by}~(\ref{E_unsat}), \\
    2.1~P_{0.1}^{-1/4} & {\rm if}~E_{\parallel}~
		{\rm is~given~by}~(\ref{E_sat}).
\end{array} 
\right.
\ee
By substituting expressions for $S_a$ and $\gamma _{min}$ into 
(\ref{S0_2}) we get 
\be 
z_0 \equiv S_0/R = 0.03 \left\{ \begin{array}{ll}
    1.9~P_{0.1}^{11/14}B_{12}^{-4/7} & {\rm if}\: E_{\parallel } ~
		{\rm is~given~by}~(\ref{E_unsat}), \\
    3.0~P_{0.1}^{7/4} B_{12}^{-1}& {\rm if}~E_{\parallel}~
		{\rm is~given~by}~(\ref{E_sat}).
\end{array} 
\right.
\label{z0_1}
\ee
Note that formulae (\ref{E_unsat}) and (\ref{E_sat}) for the 
accelerating electric field are most likely applicable when 
$z_0 < r_{pc}/R$ and $z_0 > r_{pc}/R$, respectively. Using the above 
expressions for $z_0$, these two criteria can be reduced to 
$P_{0.1}^{9/4} < 0.5~B_{12}$ and $P_{0.1}^{9/4} > 0.4~B_{12}$, 
correspondingly. Then, using relation $P_{0.1}/B_{12} \approx 1.24 
\sqrt{\tau _6}$~($\tau _6 \equiv \tau/10^6$ yr, $\tau = P/2{\dot{P}}$ 
is the pulsar spin-down age) we can rewrite formula (\ref{z0_1}) as 
\be 
z_0 \approx 0.01 \tau _6^{1/2} 
\left\{ \begin{array}{ll}
    0.7~ B_{12}^{3/7}P_{0.1}^{-3/14} & {\rm if}\: P_{0.1}^{9/4} < 0.5~B_{12}, \\
    1.1~P_{0.1}^{3/4} & {\rm if}~P_{0.1}^{9/4} > 0.4~B_{12}.
\end{array} 
\right.
\label{z0_2}
\ee
For a Crab-like PSR with $\tau \sim 10^3$ yr, $P_{0.1} = 
0.3$ and $B_{12} = 8$ the above formula yields
\be
z_0 \approx 0.007, 
\ee
whereas for middle-aged PSRs with $\tau \sim 5\cdot 10^4-10^5$ yr,  
$P_{0.1} = 2-3$, and $B_{12} = 7-8$ we get 
\be
z_0 \approx 0.01-0.02.  
\ee
Finally, for an old 10-ms PSR with $\tau \sim 10^8$ yr and $B_{12}=
8\cdot 10^{-3}$, we get 
\be
z_0 \approx 0.2 .
\ee 

Inserting Eqn. (\ref{z0_2}) into Eqn. (\ref{rho+/rhoGJ}) we arrive at 
the following expression for fractional returning positron density 
\be 
{\rho _+ \over \rho_{\sc GJ}} \approx 0.01 \tau _6^{1/2} 
\left\{ \begin{array}{ll}
    1.85~B_{12}^{3/7}P_{0.1}^{-3/14} & {\rm if}\: P_{0.1}^{9/4} < 0.5~B_{12}, \\
    3.0~P_{0.1}^{3/4} & {\rm if}~P_{0.1}^{9/4} > 0.4~B_{12}.
\end{array} 
\right.
\label{rho+/rhoGJ_2}
\ee

\subsection{Numerical calculation of the returning positron fraction}
\subsubsection{Pair source function in screening region}  \label{sec:Psf}

The pair source function $Q^{\pm}(\gamma^{\pm}_0, x_{nr})$ is the joint initial 
energy and spatial distribution of electron-positron pairs produced by the 
primary electrons, where $x_{nr}$ is the distance above $z_0~R$.  Only the
first generation of pairs is important to the screening process, as the higher pair
generations are produced beyond a distance of $z_0 + N_s\Delta_s$ 
($N_s$ is the number of screening scales specified in the end of Section 3.2.2).  
It can be shown that the attenuation length of synchrotron photons 
from the first pair generation is much larger than $\Delta_s$.  Each
primary electron is accelerated from its starting point at the stellar surface, 
radiates CR photons and the pair production attenuation lengths $L$ of these 
photons are computed, as described in detail by HM98.  The location of the 
first pair defines the PFF at $z_0$.  We compute the pair source function by 
accumulating the number of pairs per primary particle at each energy and height 
above the PFF as the particles accelerate up to the PFF, averaged over 
simulations for 10 - 50 primary electrons.  The primary particle trajectory 
is divided into discrete steps and at each step the CR spectrum at energy $\gamma(s)$
is divided into discrete energy bins.  A representative photon from each CR energy bin
is propagated through the local field to determine whether it produces a pair or
escapes (for details of such a calculation, see Harding et al. 1997).  
We accumulate a survival probability,
\be
   P_{\rm surv}(s) = \exp\Bigl\{-\tau_{\gamma}(s)\Bigr\} ,
\ee
where
\be
   \tau_{\gamma}(s) = \int_0^s T(\theta_{\rm kB}, \omega ) ds'
\ee
is the optical depth along the path.  The PC angles of ms pulsars are
large enough that the curvature of the photon trajectories in the 
strong gravitational field of the NS is important and we take this 
into account in computing the pair attentuation lengths (see Harding 
et al. 1997, for details). A random number $\Re$ is chosen to determine the 
pair production point of each test photon, when $P_{\rm surv} (s=L) = \Re$.  
The momentum of each pair member is assumed to have half the energy and the 
same momentum, parallel to the local magnetic field, as the parent photon.  
The ``number" of CR photons, $n_{\sc CR}$, represented by the test photon 
in each energy bin is estimated by dividing the energy radiated in the 
spectral interval, 
$\Delta\varepsilon^i = \varepsilon^i_{\rm max}-\varepsilon^i_{\rm min}$, 
by the average energy in that interval, 
$\langle \varepsilon ^i \rangle $,
\be
n^i_{\rm CR}(\langle \varepsilon ^i \rangle ) = {\dot\gamma_{_{\rm CR}}\Delta s 
\over {\langle \varepsilon ^i \rangle }c}
{\int_{\varepsilon^i_{\rm min}}^{\varepsilon^i_{\rm max}} 
N_{\rm CR}(\varepsilon) d\varepsilon
\over \int_0^{\infty} N_{\rm CR}(\varepsilon) d\varepsilon} ,
\ee
where
\be
N_{\rm CR}(\varepsilon) = {\sqrt{3} e^2\over c}\gamma\,
F(\varepsilon / \varepsilon_{cr})
\ee
is the CR energy spectrum, $\varepsilon_{cr}$ is the
critical energy and $\rho_c$ the field line radius of curvature 
[see also the formulae right after Eqn. (\ref{n_+/n_e}) ] and 
$\dot\gamma_{_{\rm CR}}$ is the CR loss rate 
$\dot\gamma_{_{\rm CR}}mc^2 = 2e^2 c\gamma ^4 /3 \rho _c^2$.
The height above the PFF and the energy of the pairs are accumulated 
in a two-dimensional distribution, normalized by the total number of 
test primary photons and by $n_{\sc CR}$, to form a 
distribution $Q^{\pm}(\gamma^{\pm}_0, x_{nr})$ of the number of pairs 
per energy per primary at each height and energy interval.  An
example of one of the computed pair source functions is shown in Figure 1.  
The first pairs have the highest energies because they have the shortest 
attenuation lengths.  The number of pairs increases rapidly beyond the 
PFF and the mean energy of the pairs decreases.

\subsubsection{Pair dynamics}

The charge density along open field lines above the PFF will increase 
due to injection of pairs.  The dynamical response of the pairs to the 
$E_{\parallel}^{sc}$, i.e. acceleration of electrons and deceleration 
of positrons, will determine the increase in effective space-charge 
density with height in the screening region.  We find that the 
$E_{\parallel}^{sc}$ is large enough that the major contribution to 
the space-charge density results from positrons that are
turned around and accelerated back toward the NS surface.  
Above the PFF, the continuity equations for elections and positrons are
\be   \label{cont+}
{d n^+ \over d t} = {\partial {n^+} \over \partial t} + 
{\partial (n^+ \beta^+c) \over \partial x} ,
\ee
\be   \label{cont-}
{d n^- \over d t} = {\partial {n^-} \over \partial t} + 
{\partial (n^- \beta^-c) \over \partial x} ,
\ee
where $n^+$ and $n^-$ are the first moments and $\beta^+$ and $\beta^-$ 
are the second moments of the positron and electron distribution 
functions, $f^{\pm} (\gamma^{\pm}, x)$, respectively:
\be  \label{beta}
\beta^{\pm} = {u^{\pm}\over c} = \int {v^{\pm}\over c} f^{\pm} 
(\gamma^{\pm}, x) dv
\ee
\be  \label{n}
n^{\pm} = \int f^{\pm} (\gamma^{\pm}, x) dv.
\ee
Since the source function is electron positron pairs, 
$\dot n^+ = \dot n^-$, we can subtract Eqn. (\ref{cont-}) from Eqn. 
(\ref{cont+}) to give 
\be
e{d (n^+ - n^-) \over dx} = {d \rho \over dx} = e \left(
{\partial n^+ \beta^+ \over \partial x}
- {\partial n^- \beta^- \over \partial x} \right) ,
\ee
so that the total charge density is
\be  \label{rhox}
\rho = e (n^+\beta^+ - n^-\beta^-) .
\ee
The distribution functions of electrons and positrons 
$f^{\pm} (v^{\pm}, x)$ are computed by dynamically evolving the pair 
source function, subject only to the force of the electric field in 
the screening region.  Pairs are injected at discrete points, $x_{nr}$, 
and with a discrete distribution of energies, $\gamma^{\pm}_0$ as described below. 
Each sign of charge in each of the $n_e$ energy bins is evolved in energy through a
separate grid, $x = (z - z_0)R$, of $E_{\parallel}^{sc}(x)$ computed at discrete 
points $x_n$. Using Eqn. (\ref{E_pff}), we find that
with the resolution of the calculation, the region in which the positrons are 
non-relativistic at their turn-around points is not resolved.  We are therefore 
justified in treating the particles as relativistic everywhere and may thus 
ignore the momentum equation.  The energy of a particle at point $x$ in the grid is 
\be
\gamma^{\pm}(x) = \gamma_0^{\pm} \mp {e\over mc^2} 
\int_{x_{nr}}^x E_{\parallel}^{sc}(x')dx' ,
\ee
so that
\be
{ {v^{\pm}(x)} \over c} = 
\left\{ 1 - {1\over {[\gamma^{\pm}(x)]^2}}  \right\} ^{1/2} .
\ee
Particles from the source function, $Q^{\pm}(\gamma^{\pm}_0, x_{nr})$ 
are evolved through the grid until they either
travel upward across the top boundary or downward across the lower 
boundary.  At each grid height, the steady-state distribution function 
of electrons and positrons are computed
\be \label{f}
f^{\pm} (\gamma^{\pm}, x) = \sum_{nr} 
{Q^{\pm}(\gamma^{\pm}_0, x_{nr})\over {v^{\pm}(x)}}.
\ee
Using Eqns (\ref{beta}), (\ref{n}), (\ref{rhox}) and (\ref{f}), we can 
determine the density of upward-moving electrons, $n^-_u (x)$, and 
positrons, $n^+_u (x)$ and the downward-moving positrons, 
$n^+_d (x)$ at each grid point. The total charge density due to pairs is then
\be
\rho(x) = e [n^+_u (x) \beta^+_u (x) - n^-_u (x) \beta^-_u (x) + 
n^+_d (x) \beta^+_d (x)].
\ee

The downward-moving positrons will constitute an additional upward-moving 
negative current which will add to the current of primary elections 
(Section 2.2). Because the net current $j$ is required to be the GJ current, 
we therefore need to adjust the primary current for the returning 
positron current. Thus, we can write 
\be
n^-_p (1) = n^-_p - n^+_d (1),
\ee
where $n^-_p$ is the number density of primary electrons and
$n^+_d (1)$ is the downward-moving positron density at the 
grid lower boundary (the PFF).  
The charge deficit, $\Delta\rho = \rho^-_p - \rho_{GJ}$ at the 
PFF is thus also readjusted. 

The calculation thus proceeds as follows.  We first compute the location 
of the PFF above the surface, $z_0$, and the pair source functions as 
described in Section \ref{sec:Psf}. An initial screening scale length, 
$\Delta_s$, is chosen as a fraction of $z_0$, the distance to the PFF.  
The electric field in the screening region, above the PFF, is modeled by 
an exponential (Eqn. [\ref{E_pff}]) with scale length $\Delta_s$. 
The pair dynamics determines a pair distribution function and a 
returning positron density, as described above.  The primary electron 
flux and $\Delta\rho$ are adjusted for this returning positron flux, 
and the pair distribution function and a returning positron density are 
then recomputed using this adjusted charge deficit.  A new screening 
scale length is determined from the point where $\rho(N_s\Delta_s) = 
\Delta\rho$, where $N_s$ is the number of $E_{\parallel}^{sc}(x)$
scale heights required to guarantee that no positrons turn around above 
$z_0 + N_s\Delta_s$. The solution will then be self-consistent.  
We find that $N_s = 4.0$ satisfies this condition and set it as a 
constant value.  The scale height of the electric field in the 
screening region is then set to the new $\Delta_s$.  The iteration 
continues until the returning positron density and screening scale 
attain stable values.

\subsubsection{Screening scale height and returning positron density}

Convergence of the screening scale height $\Delta_s$, the total charge 
density above the PFF, $\rho(x)/\rho_{GJ}$, and the relative returning 
positron density, $\rho^+_d(1)$, is achieved independently at each 
colatitude, $\xi$, across the PC.  An example of a self-consistent 
solution for $\rho(x)/\rho_{GJ}$ and $\Delta_s$ is shown in Figure 2.  
The charge density increases rapidly immediately above the PFF, 
where the pair source function is growing exponentially and where 
$E_{\parallel}^{sc}$ is high enough to decelerate and turn around 
all the produced positrons. At a distance above the PFF comparable 
to the screening scale height, $E_{\parallel}^{sc}$ has decreased 
to a fraction of its value at the PFF and the rate of increase of
the charge density begins to moderate, as fewer positrons are able 
to turn around. We find that the creation of the charge density and 
subsequent screening of $E_{\parallel}^{sc}$ is due almost entirely 
to returned positrons rather than to velocity differences between 
electrons and positrons as the pairs dynamically respond to the 
electric field above the PFF.  Due to the boundary condition imposed 
on the potential (see Eqn. [\ref{Phi_pff}]), $E_{\parallel}^{sc}$ 
never decreases to 0, but only approaches 0 at infinity.  
Consequently, the self-consistent charge density never achieves 
the value $\Delta\rho$ but approaches it asympotically from below. 

Figure 3 shows solutions for the screening scale, $\Delta_s$, and the returning
positron density, $\rho_+/\rho_{GJ}$, across the PC as a function of the
colatitude $\xi$, scaled to the PC opening angle at the surface, for various 
values of the pulsar period and surface magnetic field strength.  Both $\Delta_s$
and $\rho_+/\rho_{GJ}$ increase toward $\xi = 0$ (magnetic pole) and $\xi = 1$
(the outer edge of the PC).  Since the value of the GJ density,
$\rho_{GJ}$ is constant with $\xi$, the variation represents a true variation in
$\rho_+$ across the PC.  The increase in $\Delta_s$ and $\rho_+$ toward $\xi = 0$ 
is due to the increasing field line radius of curvature near the pole, causing the 
pair attenuation length to grow large.  The pair source function grows more slowly
with distance above the PFF, and the screening scale height increases.  The returning
positron density increases because as the height of the PFF increases, the charge
deficit also increases, requiring more returning positrons for screening. 
Near the pole, there are no solutions for $\Delta_s$ and $\rho_+$ and screening 
is either incomplete or there is no screening at all.  The increase in $\Delta_s$ 
and $\rho_+$ toward $\xi = 1$ is due to the decrease in $E_{\parallel}$ caused by 
the boundary condition $\Phi = 0$ at $\xi = 1$ imposed on the solution to Poisson's 
equation.  Consequently, the primary particles must accelerate over a longer 
distance to produce pairs, increasing the height of the PFF
and thus $\rho_+$.  Also the positron turn-around distance becomes longer with a
lower $E_{\parallel}$, increasing the screening scale height.

However, the variation in $\Delta_s$ and $\rho_+(\xi)$ with $\xi$ is small compared to 
their variation with pulsar period and surface field strength.  Comparing Figures
3b,3c, and 3d, one can see that the screening scale height $\Delta_s$ increases by a
factor of about ten as the period increases by only a factor of four, for a
constant surface field.  Comparing Figures 3a and 3c at a period of 0.1 s, $\Delta_s$
decreases by a factor of about six as the surface field increases by only a factor of two. 
Figure 4 shows the dependence of the screening scale length, $\Delta_s$, on surface field 
and period.  Generally, $\Delta_s$ decreases with increasing surface field strength
because the pair attenuation length is shorter in higher fields.  As a result, the pair
density grows faster and $E_{\parallel}$ can be screened in a shorter distance above
the PFF.  Consequently, ms pulsars have relatively large screening scale
lengths, but $\Delta_s$ is still small compared to $z_0$. 
Figure 5 shows the dependence of $\rho_+$ on surface field and period.  While $\rho_+$
is less sensitive to pulsar parameters than $\Delta_s$, it decreases significantly
as the surface field increases and as the period decreases.  The numerical values of
$\rho_+/\rho_{\rm GJ}$ are smaller than, but within a factor of two of, the analytic 
estimate of Eqn. (\ref{rho+/rhoGJ_2}). The variation of 
$\Delta_s$ and $\rho_+$ with pulsar parameters is primarily due to the dependence of
$E_{\parallel}$ and the pair production attenuation length on period and surface 
field strength. By contrast, $\Delta_s$ and $\rho_+(\xi)$ vary by only factors of two 
or three across the PC of a single pulsar. 

Since $\rho_+$ is always a small fraction of both the primary particle density and of
$\rho_{\rm GJ}$, only the very first pairs of the full, multi-generation cascade are
needed to screen the $E_\parallel$.  We therefore did not simulate the full
cascade in computing the pair source functions needed in the present work. 
The vast majority of the cascade pairs are produced
in the region where $E_\parallel \sim 0$ and they freely escape the magnetosphere 
with only radiation losses, but no significant acceleration.  Previous simulations 
of the full cascade produced by the primary particles (Daugherty \& Harding 1996) give 
multiplicities $\sim 10^3 - 10^4$ pairs/primary.  The fraction of the total number of
secondary pairs that are returned to heat the stellar surface can thus be estimated
by dividing the values of $\rho_+/\rho_{\rm GJ}$ in Figure 5 by the cascade 
multiplicity.  

\section{POLAR CAP HEATING RATES AND THERMAL X-RAY LUMINOSITY}
\subsection{Analytic estimate}

Let us derive an expression for the total power that can be deposited 
onto a single PC by precipitating positrons. The general expression 
can be written as 
\be
L_+ = \alpha c \int _{{\cal S}(z_0)} \rho _+(z_0,\xi) \Phi (z_0,\xi) dS,
\label{L_+I}
\ee
where the integration is over the area of a sphere cut by the polar flux tube 
at the radial distance $\eta _0$, and the factor $\alpha $ accounts for the 
general-relativistic correction to the current ($j \sim \alpha c \rho $). 
Here $dS = [S_{pc} \eta _0^3 f(1)/\pi f(\eta _0)] d \Omega _{\xi }$, 
$S_{pc} = \pi \Omega R^3 /c f(1)$ is the area of the PC, $d\Omega _{\xi } = 
\xi d\xi d\phi $ is an element of a solid angle in the PC region. 
Thus expression (\ref{L_+I}) reduces to 
\be
L_+ = 2 \alpha c S_{pc} \eta _0^3 {{f(1)}
\over {f(\eta _0)}} \int _0^1 \rho _+(z_0,\xi) \Phi (z_0,\xi )\xi d\xi.
\label{L_+II}
\ee
After inserting expression for $\Phi $ [see Eqns (\ref{Phi_unsat}), 
(\ref{Phi_sat})] into (\ref{L_+II}) and normalizing $\rho _+$ by 
$\rho _{\sc {GJ}}$, we get
\be
L_+ = f_+ {\dot E}_{rot},
\label{L_+III}
\ee
where
\be
f_+ = {10 \over {\theta _0 \sqrt{1-\epsilon}}} \kappa 
(1-\kappa /\eta _0^3) z_0^2 \cos ^2 \chi  \int _0^1 
{{\rho _+(z_0,\xi)}\over {\rho_{\sc {GJ}}(z_0)}}
(1-\xi ^{2.19})^{0.705} \xi d\xi 
\label{f_+unsat}
\ee 
and
\be
f_+ = 9 \kappa (1-\kappa /\eta _0^3) z_0 \cos ^2 \chi  \int _0^1 
{{\rho _+(z_0,\xi)}\over {\rho_{\sc {GJ}}(z_0)}}(1-\xi ^2)\xi d\xi 
\label{f_+sat}
\ee 
are the fractions of pulsar spin-down power consumed by returning 
positrons, corresponding to the cases where $\Phi $ is given by 
Eqn. (\ref{Phi_unsat}) and (\ref{Phi_sat}), respectively.  
Here 
\be
{\dot E}_{rot} = {1\over 6} {{\Omega ^4 B_0^2R^6}\over {c^3f^2(1)}}
\label{E_rot}
\ee
is the general-relativistic expression for the pulsar spin-down losses 
in vacuum (see MH97). In this formula $B_0/f(1)$ is the surface value 
of the magnetic field strength ``as seen" by an infinitely remote 
observer, whereas $B_0$ is the value measured locally, at the NS surface. 
In a flat-space limit $B_0/f(1)$ simply transforms into $B_0$, and we 
get a classical formula for the magneto-dipole losses. Note that in our 
derivation of formulae (\ref{f_+unsat}) and (\ref{f_+sat}) we used only 
the components of potential and charge density that are proportional 
to $\cos \chi $ and we assumed that $z_0$ is independent of $\xi $ (flat PFFs), 
which is still a satisfactory approximation for rough analytic estimates. 
Our numerical calculations of $f_+$ take into account the 
curvature of PFFs (see further discussion of the numerical results), though.

Now, using expression (\ref{z0_2}) for $z_0$ and (\ref{rho+/rhoGJ_2}) for 
$\rho _+/\rho _{\sc GJ}$, we can combine Eqns (\ref{f_+unsat}) and 
(\ref{f_+sat}) and write (setting $\cos\chi \simeq 1, \kappa \simeq 0.15$)
\be
f_+ \approx 10^{-3} \tau_6 
\left\{ \begin{array}{ll}
    0.88~ (P_{0.1}^8/\tau _6 )^{1/7} & {\rm if}\: P_{0.1}^{9/4} < 0.5~B_{12}, \\
    0.96~P_{0.1}^{3/2} & {\rm if}~P_{0.1}^{9/4} > 0.4~B_{12}.
\end{array} 
\right.
\label{f_+}
\ee
This expression illustrates that there is a rather strong correlation 
between the efficiencies of PC heating by precipitating 
positrons and pulsar spin-down age. The important implication of this 
correlation is that older pulsars should favour enhanced PC heating
(see, however, the warning on use of Eqn. [\ref{f_+}] in Section
\ref{sec:numL+}). It is interesting that the above order-of-magnitude 
analytic estimates are in good agreement with our numerical calculation.

A81 derived an expression for PC heating luminosity, taking
into account the curvature of field lines but not general-relativistic
effects in the derivation of $E_{\parallel }$. Using his results 
we can derive the following expressions for $f_+$
\be
f_+^A \approx 10^{-4} \tau _6
\left\{ \begin{array}{ll}
    0.7~(P_{0.1}^3/\tau _6)^{1/5} & {\rm if}\: P_{0.1} < 3, \\
    0.05~P_{0.1}^{-3/8}\tau _6^{-1/2} & {\rm if}~P_{0.1}> 3.
\end{array} 
\right.
\label{Aronsf_+}
\ee
The value of $f_+^A$ in Arons' model also increases with pulsar age 
but is significantly lower than our result.  This is because the 
frame dragging electric field and consequent acceleration energy 
in our model is higher.

During the photon cooling era, the effective temperature of a NS 
decreases with age according to a power law. For example, using 
the cooling calculations by Page and Sarmiento (1996), we can 
estimate that the cooling luminosity, $L_{cool}$, roughly scales as 
$\tau _6^{-4}$ ($\tau _6 \geq 0.3$) and $\tau _6^{-3}$ 
($\tau _6 \geq 1$) for the case with and without core 
neutron $^3P_2$ pairing, respectively. By normalizing the 
luminosity of a cooling NS by ${\dot E}_{rot}$, we get the 
following expression for the cooling efficiency
\be
f_{cool} = {L_{cool}\over {\dot E}_{rot}} \approx 10^{-4} \tau _6
\left\{ \begin{array}{ll}
    0.15~(P_{0.1}/\tau _6^2)^2 & {\rm if}\: 
		{\rm core~neutron~} ^3P_2- {\rm pairing~and~} \tau _6 \geq 0.3, \\
    2.5~P_{0.1}^2/\tau_6^3 & {\rm if}~
		{\rm no~core~neutron~} ^3P_2-{\rm pairing~and~} \tau \geq 1.
\end{array}
\right.
\label{f_cool}
\ee
>From expressions (\ref{f_+}) and (\ref{f_cool}) one can see that the 
pulsars with ages $\geq 10^6$ yr are most likely candidates for those 
with luminosities dominated by the PC heating.

\subsection{Numerical calculation of polar cap heating luminosity} \label{sec:numL+}

The returning positrons accelerate through the same potential drop as the primary 
electrons and their energy heats the PC surface.  We can calculate the luminosity 
heating the PC by using expression (\ref{L_+II}) and substituting the corresponding 
quantities from the numerical calculation. Figure 6 shows the dependence of 
$L_+ (\xi)$ on $\xi$ (from the integrand of Eqn. [\ref{L_+II}]), which reflects 
the distribution of heating across the PC. Although the distribution of 
$\rho_+(z_0,\xi)$ has maxima at $\xi = 0$ and $\xi = 1$ (see Figure 3), 
$\gamma_{\rm max}(\xi)$ has a strong maximum at $\xi = 0$ and 
decreases monotonically with $\xi$ (see Figure 4 and 5 of HM98), 
causing $L_+$ to decrease with $\xi$.  

The total positron heating luminosity, scaled with the spin-down luminosity as a 
function of characteristic pulsar age, $P/2\dot P$ is shown in Figure 7.  The
numerically computed $L_+/\dot E_{rot}$ increases nearly linearly with $\tau $
in the same way as the analytic estimate in the unsaturated regime (first expression
of Eqn. [\ref{f_+}]).  $L_+/\dot E_{rot}$ also increases approximately linearly 
with period for a constant age.  The PC heating luminosity is thus a negligible 
fraction of the spin-down luminosity in young pulsars with ages less than 
$\tau \sim 10^5$ yr, but becomes a significant source of heating in older pulsars. 
The numerical values are a factor of $\sim 6$ lower than the estimated
values of $L_+/\dot E_{rot}$ from the first expression of Eqn. (\ref{f_+}) 
for the normal pulsars.  For the ms pulsars, the analytic values are somewhat 
higher than the numerical values, but by less than a factor of ten.  The reason 
that the analytic estimate, which simply assumes that the returning positron 
flux is half of the charge deficit at the PFF, is closer to the numerical 
values for normal-period pulsars may be that the screening scale heights 
are small and the returning positron fraction is not very sensitive to the 
structure of the electric field in the screening region.  On the
other hand the screening scale heights for the ms pulsars are about an order
of magnitude larger, reflecting the larger (see Figure 4) growth scale of 
the pair source function. Consequently, the returning positron fraction 
for ms pulsars is more sensitive to the structure of the electric field 
in the screening region. 

Although the analytic estimate of Eqn. (\ref{f_+}) seems to be a very good estimate
of the returning positron luminosity, we caution that it cannot be extrapolated to
ages much beyond the numerical results, i.e. above $\tau \sim 10^7$ yr for normal pulsars
and above $\tau \sim 10^8$ yr for ms pulsars. This is because the assumption
of complete screening breaks down at age-period combinations where the pulsar cannot
produce enough pairs.  When the pair density grows too slowly and cannot reach a high enough
level for complete screening, then the returning positron fraction drops below the
charge deficit prediction.  Beyond the upper end of each constant period line in Figure 7,
numerical solutions of returning positron fraction and screening scale length do not exist.
In fact, complete CR screening terminates on the constant period line at about the same 
point where PFFs are no longer produced.  Beyond this point, the returning positron
fraction and the heating luminosity will drop sharply.  However, as we will discuss in
our next paper, ICS pair fronts are produced by older pulsars, even when the
ICS screening from these pairs is incomplete.  This will result in a PC heating
component due to returning ICS positrons from partial screening which can be quite 
significant for ms pulsars.  

The results in Figure 7 have assumed a constant value of inclination angle, 
$\chi = 0.5$. The expected variation of $f_+ \propto \cos^p\chi$ (where $p = 5/7$ or 
1/2) will cause $f_+$ ($= L_+/\dot E_{rot}$) to gradually decrease with 
increasing $\chi$ until $\chi$ approaches
$\pi/2$.  At this point, the second term in Eqn. (\ref{Phi_gen}) for the electrostatic 
potential (the relativistic equivalent of A83 solution) becomes important.  
For $\chi = \pi/2$, our solution for the PC heating rate will be comparable to 
that of Eqn. (\ref{Aronsf_+}).   

X-ray emission at energies 0.1 - 2 keV has been detected from several dozen pulsars 
by ROSAT (Becker \& Trumper 1997).  There is a rough empirical correlation of X-ray 
luminosity with spin-down luminosity, giving $L_x \sim 10^{-3} \dot E_{rot}$, so that the
observed level of X-ray emission in pulsars would be a constant line in Figure 7 at
$10^{-3}$.  The maximum calculated values of $L_+/\dot E_{rot}$ do reach the observed
level, indicating the emission from PC heating in normal pulsars with 
$\tau > 10^5$ yr and in ms pulsars is detectable.  

In Table 1, we compare our computed values of flux and surface temperature from PC 
heating with measured values of some older pulsars in which hot thermal
components have been detected. Our values are those at the NS surface and have not 
been corrected for gravitational redshift effects - i.e. $L_+$ and $\Phi_+$ would be 
about $40\%$ lower (a factor of $\alpha^2$) and $T_+$ would be about $20\%$ lower (a factor
of $\alpha$) for observers at infinity.  Greiveldinger et al.(1996) have fit 
three-component (two thermal and one power-law) spectra to combined ROSAT and ASCA 
data from PSR0656+14 and PSR1055-52.  These pulsars are young
enough to have expected cooling as well as heating thermal components, so it is important
to separate the emission from these two components in order to make comparisons with
PC heating models.  PSR0656+14, however, has an inferred surface dipole field 
of $9 \times 10^{12}$ G and so is not in the regime of CR pair fronts near the NS surface
that we have treated in this paper.  It may still have a CR pair front at higher altitude
if the ICS pair front is unstable, a situation we will address in our next paper.  
The fits for the heated area $A$ of the hot components in both cases are much smaller than 
the NS surface area, and are thus consistent with emission from heated PCs.  
Our predicted values of PC temperature from returning positrons agree fairly well 
with the measured values for these pulsars, although our predicted fluxes are significantly
lower than those observed.  Because these middle-aged pulsars are still dominated by cooling 
components and have power law components at high energies, extracting the relatively small 
heating component is very difficult.  

Wang \& Halpern (1997) have fit single-component
blackbody spectra to ASCA data of PSR1929+10.  This pulsar is too old
to have detectable cooling component and the one-component fits of ASCA 
data indeed indicate emission from heated PC whose heated area is much smaller than even
the standard PC area.  We have thus assumed the measured value of area $A$ in 
computing the theoretical PC temperatures, $T_+ = (L_+/\sigma A)^{1/4}$, of
these pulsars, instead of the canonical PC area, $A_{\rm pc} = \pi R (\Omega R/c)$.  
Our computed temperatures are in agreement with the measured values,
within the uncertainties.  Long period (older) pulsars
have smaller PCs and thus the returning positron luminosity will heat
a smaller area and we therefore would predict that pulsars with longer periods should
have higher PC temperatures. PSR1929+10 does in fact have higher measured temperature 
and smaller heated area. Although Geminga is expected to have both cooling and heating 
components, it is more difficult
to compare our theoretical results for positron heating with the observed values from
a single-component thermal fit.  The fact that our predicted flux is somewhat lower 
than the measured value is consistent with an additional cooling component.  We have 
used the canonical PC area to compute our predicted temperature for Geminga, so we would
expect it to be higher than a measured temperature which includes cooling from the whole
NS surface. 

\section{SUMMARY AND CONCLUSIONS}

In this paper we derived some practical formulae for the positron fluxes 
returning from the upper PFF in rotation-powered pulsars in cases where 
$E_{\parallel }$ screening is produced by pairs from CR. We presented 
the expected theoretical values for the PC X-ray luminosities 
due to the heating by precipitating positrons. The calculated efficiencies 
of PC heating explicitly depend on the pulsar spin-down age and 
spin period, and can be used in the analyses of thermal X-ray fluxes 
from pulsars. Our numerical calculation of returning positron flux 
and PC heating efficiencies show that the dependence on $\tau$ 
and $P$ is the same as in analytic formulae, the actual values of the 
numerical quantities are smaller than the analytic estimates by a 
factor of 2-3. In summary, we have reached the following conclusions. 

\begin{itemize}

\item The heating of the PC by returning positrons is possible if stable pair 
fronts develop, and $E_{\parallel }$ is not assumed to be fully screened at the onset 
of the PFF. In most cases we discuss in this paper the fraction of returning 
positrons is not affected by the details of our modeling of the screening of 
$E_{\parallel }$ and is mostly determined by the distribution of pairs beyond the PFF. 

\item In contrast to the results obtained earlier by A81, we find that 
the returning positrons may significantly contribute to the PC heating and 
therefore to the observed pulsar thermal X-ray fluxes, especially for older 
pulsars. 

\item Our theoretical model can be thoroughly tested against observations as soon as 
more data on the pulsed thermal X-ray fluxes from middle-aged and old 
pulsars become available. We anticipate that the pulsar X-ray light curve 
would have larger pulse fraction in the case of NS thermal emission produced by 
the PC heating than in the case of thermal emission during NS photon-era cooling. 

\item We predict that long-period pulsars will have higher surface temperatures from
PC heating than those of short-period pulsars.   

\item Our calculations indicate that the PC heating increases 
toward the magnetic axis, which could manifest itself in a fine structure of 
the X-ray light curves and heated areas smaller than the PC area.  
  
\end{itemize}

We caution that our analytic expression in Eqn. (\ref{f_+}) for $f_+$ is not applicable for 
normal pulsars with $\tau \gsim 10^7$ yr or ms pulsars with $\tau \gsim 10^8$ yr. The reason 
is that the analytic formulae are based on constant magnetic field and do not take 
into account the cessation of pair formation by CR photons beyond a certain age (dependent
on period). Thus, our plots (see Figure 7) for pulsars with 5 and 10 ms periods cannot be 
compared with the observational data for actual ms pulsars, most of which have ages 
$\tau \gsim 10^8$ yr where CR photons cannot produce pair fronts.  Although our analytic 
formula for $f_+$ would give very large heating luminosities for these ms pulsars
(beyond the region of validity of the formula), the actual heating efficiencies will
be much lower.  However, as we will address in our next study, ms pulsars
are capable of producing pairs via non-resonant ICS and the resulting heating luminosities
will be detectable.  Furthermore, the results presented in this paper are not applicable to
pulsars with surface magnetic fields above about $4 \times 10^{12}$ G, which are capable of 
ICS screening.  The possibility of ICS pairs from those pulsars which are beyond 
their CR deathlines, as well as ICS screening and PC heating in higher-field pulsar, 
will be investigated in our next paper.  

\acknowledgments 
We thank the anonymous referee for careful reading of the 
manuscript and helpful comments.  This work was supported 
by the NASA Astrophysics Theory Program.

\clearpage 

\section*{APPENDIX\\
SOLUTIONS WITH EXPLICIT BOUNDARY CONDITIONS AT THE UPPER BOUNDARY} 

Here we present the solutions to Eqn. (1) subject to the explicit boundary 
conditions (especially at the upper boundary) different from those discussed 
in the main text. These solutions illustrate how the onset of the 
upper boundary affects the entire distribution of electrostatic potential 
and accelerating electric field above the PC. It is also instructive to compare 
these solutions with those we used in our calculations (see the main 
text for details). Throughout this Section we denote the dimensionless 
height of the upper boundary by $z_c$ ($= \eta _c -1$). We 
factorize the expressions for the electrostatic potential and field 
by $\Phi _0 \equiv (\Omega R /c)B_0 R$ and $E_0 \equiv \Phi _0/R$, 
respectively. Finally, the solutions we present here imply  
the {\it space-charge limitation of current}. 

The solution for radial function ${\cal F}_i$ [see Eqns (\ref{Phi_gen})-(\ref{B_i})] for 
which the following Dirichlet boundary conditions
$$
\Phi (z=0) = \Phi (z=z_c) = 0
$$ 
are satisfied, is 
$$
{\cal F}_i = \gamma _i \left[ z - z_c {{\sinh ^2 (\gamma _i z /2)} 
\over {\sinh ^2(\gamma _i z_c/2)}} \right] -
{{\sinh [\gamma _i (z_c-z)] + \sinh(\gamma _i z) - 
\sinh(\gamma _i z_c)}\over {1-\cosh (\gamma _i z_c)}}.
\eqno (A1)
$$
Let us also present some asymptotic expressions. For the case 
where $z < z_c << r_{pc} / R$ we get
$$   
\Phi = {1\over 2} \Phi _0 {z_c \over {1-\epsilon }} 
z^2 \left( 1 - {{z^2}\over {z_c^2}}\right)
\left[ \kappa \cos \chi + {1\over 2}\theta _0 \xi H(1) \delta (1) \sin \chi 
\cos \phi \right] , 
\eqno (A2)
$$
$$
E_{\parallel } = - E_0 {z_c \over {1-\epsilon }} 
z \left( 1 - 2 {{z^2}\over {z_c^2}}\right)
\left[ \kappa \cos \chi + {1\over 2}\theta _0 \xi H(1) \delta (1) 
\sin \chi \cos \phi \right] . 
\eqno (A3)
$$
For $z << r_{pc}/R < z_c$ we arrive at formulae (\ref{Phi_unsat}), (\ref{E_unsat}). 
Finally, for $z \gsim r_{pc} / R$ we get
$$
\Phi = {3\over 2} \Phi _0 {{\Omega R}\over c} 
{z_c \over f(1)} \left\{ \kappa \left[ (1-\xi ^2) {z\over z_c} -
8 \sum _{i=1}^{\infty } {{J_0(k_i\xi )}\over {k_i^3J_1(k_i)}} 
\exp [-\gamma _i (z_c-z)] \right] \cos \chi + \right.
$$
$$ 
\left. {1\over 4}\theta _0 \xi H(1) \delta (1) \left[ 
  \xi (1-\xi ^2) {z\over z_c} -
16 \sum _{i=1}^{\infty } {{J_1({\tilde k}_i\xi )}\over {{\tilde k}_i^3J_2({\tilde k}_i)}} 
\exp [- {\tilde {\gamma }}_i (z_c-z)]
\right] \sin \chi \cos \phi ] \right\}, 
\eqno (A4)
$$
$$
E_{\parallel } = - {3\over 2} E_0 {{\Omega R}\over c}
{1\over f(1)} (1-\xi ^2)\left[ \kappa \cos \chi + 
{1\over 4}\theta _0 \xi H(1) \delta (1) \sin \chi \cos \phi \right] . 
\eqno (A5)
$$
Solution satisfying the Neumann boundary conditions 
$E_{\parallel }(z=0) = E_{\parallel }(z=z_c) = 0$ can be derived  
from the following radial function
$$
{\cal F}_i = [\gamma (\eta \eta _c - 2/\gamma ^2) \cosh \gamma 
(z_c - z) + (2\eta _c - \eta )\sinh \gamma (z_c - z) + (2 - \eta ) 
\sinh (\gamma z) -
$$
$$
\gamma (\eta - 2/\gamma ^2) \cosh (\gamma z) - (\gamma \eta _c - 
2/\gamma ) \cosh (\gamma z_c) - (2\eta _c -1) \sinh (\gamma z_c) + \gamma -2/\gamma + 
$$
$$
(\gamma ^2 \eta _c -1)z \sinh (\gamma z_c) + 
\gamma z_c z \cosh (\gamma z_c) ] /[\gamma \eta _c 
\sinh (\gamma z_c) - \cosh (\gamma z_c) + 1].
\eqno (A6)
$$
For $z_c << 0.1(r_{\rm pc}/R)^2$ the solution can be approximated by 
$$
\Phi = {3\over 2}\Phi _0 {z_c \over {1-\epsilon }} 
z^2 \left( 1 - {2z \over 3z_c}\right)
\left[ \kappa \cos \chi + {1\over 2}\theta _0 \xi H(1) 
\delta (1) \sin \chi \cos \phi \right],  
\eqno (A7)
$$
$$
E_{\parallel } = - 3 E_0 {z_c \over {1-\epsilon }} 
z \left( 1 - {z\over z_c}\right)
\left[ \kappa \cos \chi + {1\over 2}\theta _0 \xi H(1) 
\delta (1) \sin \chi \cos \phi \right] .  
\eqno (A8)
$$
For $z < 0.3 r_{\rm pc}/R < z_c$  we again arrive at formulae (\ref{Phi_unsat}), (\ref{E_unsat}), 
whereas for $z_c >> 0.1(r_{\rm pc}/R)^2$ the solution reduces to 
$$
\Phi = {9\over 4} \Phi _0 {{\Omega R}\over c} 
{z \over f(1)} {z\over z_c}\left( 1- {{2z} \over {3z_c}} 
\right) (1-\xi ^2) \left[ \kappa \cos \chi + {1\over 4}\theta _0 \xi H(1) 
\delta (1) \sin \chi \cos \phi \right] , 
\eqno (A9)
$$
$$
E_{\parallel } = - {9\over 2} E_0 {{\Omega R}\over c} 
{1\over f(1)} {z\over z_c}\left( 1-{z\over z_c} \right) (1-\xi ^2)
\left[ \kappa \cos \chi + {1\over 4}\theta _0 \xi H(1) 
\delta (1) \sin \chi \cos \phi \right] . 
\eqno (A10)
$$ 
In the above equations the correction factors $f$, $H$, and $\delta $ 
accounting for the gravitational redshift effect read (see also HM98)
$$
f(x) = - 3\left( {x \over \epsilon } \right) ^3 
\left[ \ln \left( 1 - {\epsilon \over x} \right) + 
{\epsilon \over x } \left( 1 + {\epsilon \over {2x}} \right) \right] ,
\eqno (A11)
$$ 
$$
H(x) = {\epsilon \over x} - {\kappa \over x^3} + 
{{1 - 3 \epsilon /2x + \kappa /2x^3}\over {(1 - \epsilon /x)f(x)} } ,
\eqno (A12)
$$ 
$$
\delta (x) = \partial \ln [ H(x) \theta (x) ] / \partial x ,
\eqno (A13)
$$ 
where $x \geq 1$, and $\theta (x)$ is the half-opening angle of the 
polar magnetic flux tube defined right after Eqn. (3).

Note that in the case $z_c << r_{pc}/R$ the imposing of the boundary 
condition $\Phi (z_c) = 0$ additionally supresses the electric field 
near the stellar surface [cf. Eqns (A3) and (A8)] compared to the case  
with the boundary condition $E_{\parallel }(z_c) = 0$. This is a clear 
illustration of the fact that $\Phi (z_c) = 0$ condition forces 
$E_{\parallel }$ to vanish well below $z_c$. This effect is especially 
pronounced when the upper boundary is very close to the stellar surface 
(e.g. when $z_c << r_{pc}/R$). 

By comparing expressions (A2), (A3) with (A7)-(A10) one can see that the 
altitudinal offset between the vanishing of electrostatic potential and 
vanishing of parallel component of the electric field ranges from 0.3 to 
0.5 $z_c$. This supports our conclusion (see \S 2.3) that 
simultaneous vanishing of both $E_{\parallel }$ 
and $\nabla _{\parallel }\cdot E_{\parallel }$ hardly occurs within the 
PFF. Note also that Eqns (A5) and (A8) are the same as Eqns (A5) and (A1), 
respectively, presented in HM98. 

Recently, Dyks \& Rudak (2000) explored some useful approximations to
$E_\parallel $ that correspond to our general solution (A6). In 
particular, they arrive at the same expressions as (A8) and (19) 
[see their Eqns (9) and (12), respectively]. Here it is worth 
mentioning that in our previous paper (HM98) one of the approximate 
formulae, Eqn. (A3), is erroneous and should be replaced by e.g. 
formula (21) of this paper or by similar fitting formula given 
by Dyks \& Rudak. Additional study aimed at the fitting of numerous
combersome analytic expressions by simple compact formulae would be
desirable.


\newpage

\begin{table}
\centerline{Heated Pulsar Polar Caps}
\begin{center}
\begin{tabular}{lcccc}
      & 0656+14 & 1055-52 & 1929+10 & Geminga \\
\hline
P (s) & 0.384   & 0.197   & 0.226   & 0.237   \\
$\tau$ (yr) & $1.1 \times 10^5$ & $5.6 \times 10^5$ & $3.2 \times 10^6$ & $3.4 \times 10^5$ \\
\hline
$\Phi_{\rm h}$ (erg cm$^{-2}$ s$^{-1}$) &
       $2.4^{+1.7}_{-1.2}\times 10^{-12}$ & $1.0^{+0.7}_{-0.3}\times 10^{-13}$ &
       $1.7 \times 10^{-13}$ & $4.78 \times 10^{-12}$ \\
$A$ (cm$^{2}$) &
      $2.5^{+7.9}_{-1.7}\times 10^{11}$ & $1.1^{+8.2}_{-1.0}\times 10^{9}$ &
      $3 \times 10^{7}$ & \\
$T_{\rm h}$ (K) &
      $1.5^{+0.2}_{-0.2}  \times 10^{6}$ & $3.7^{+2.0}_{-1.2}\times 10^{6}$ &
      $5.14^{+0.53}_{-0.53}\times 10^{6}$ & $5.6^{+0.6}_{-0.6}\times 10^{5}$ \\
\hline
$L_+$ (erg/s)  &
      $4.6 \times 10^{30}$ & $5 \times 10^{30}$ & $4 \times 10^{30}$ & $6.8 \times 10^{30}$ \\
$\Phi_+$ (erg cm$^{-2}$ s$^{-1}$) &
      $1.5 \times 10^{-13}$ & $2.3 \times 10^{-14}$ & $5.6 \times 10^{-13}$ & $2.3 \times 10^{-12}$ \\
$T_+$ (K)  &
      $2.6 \times 10^{6}$ & $2.3 \times 10^{6}$ & $6.9 \times 10^{6}$ & $2.5 \times 10^{6}$ \\
\hline
\\
\end{tabular}
\caption{Measured values of hot thermal component flux $\Phi_{\rm h}$, 
heated area $A$ and temperature $T_{\rm h}$ of PSR0656+14 and PSR1055-52 are from 
Greiveldinger et al.(1996) and of PSR1929+10 is from Wang \& Halpern (1997).
Measured values for Geminga are for the total thermal component (Halpern
\& Wang 1997). In computing $\Phi_+$ from $L_+$, we have assumed a solid angle of $4\pi$.}   
\end{center} 
\end{table}

\newpage

\figureout{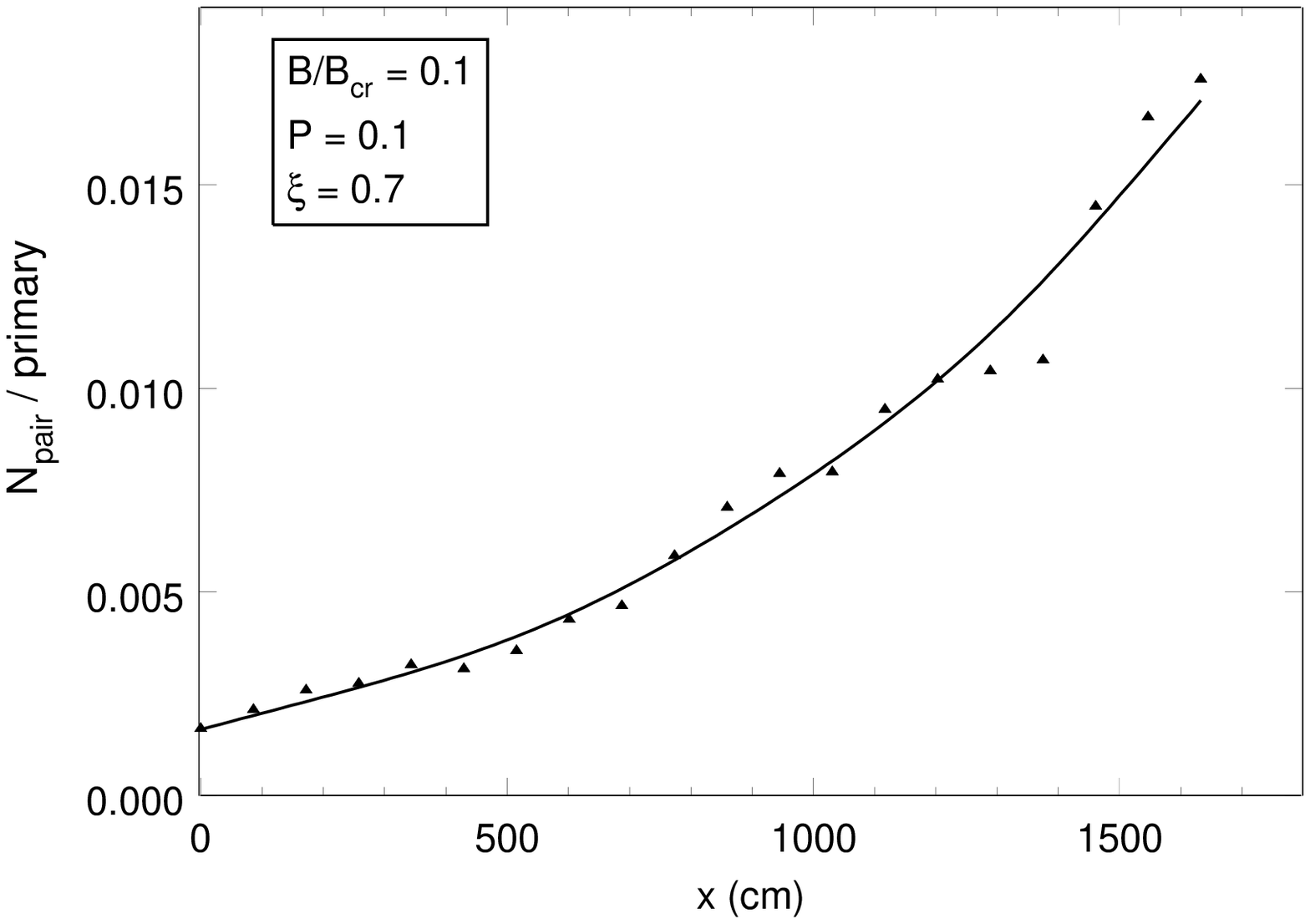}{0}{
Example of pair source function, integrated over energy, as a function of distance, x,
above the PFF.  The vertical axis measures the number of pairs
produced by CR photons in each spatial bin, normalized per primary electron.
    \label{fig:pair_source} }    

\figureout{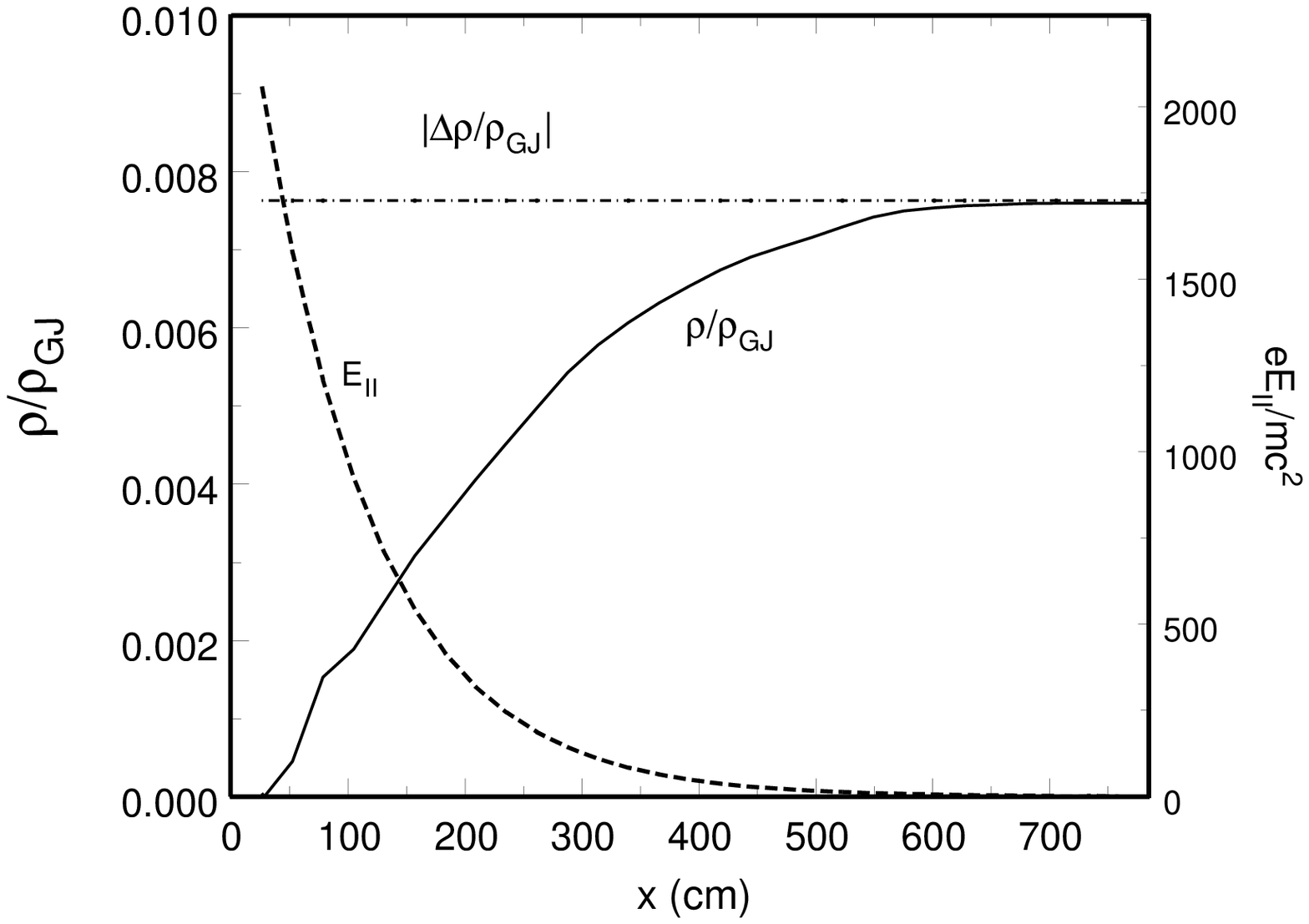}{0}{
Self-consistent solution for the charge density, $\rho$, due to returning 
positrons which asymptotically approaches the charge deficit, $\Delta\rho$, 
needed to screen the electric field, $E_{\parallel}$, above the PFF, modeled 
as a declining exponential with screening scale height, $\Delta_s R$.  
The pulsar parameters are the same as those of Figure 1. 
    \label{fig:} }    

\figureout{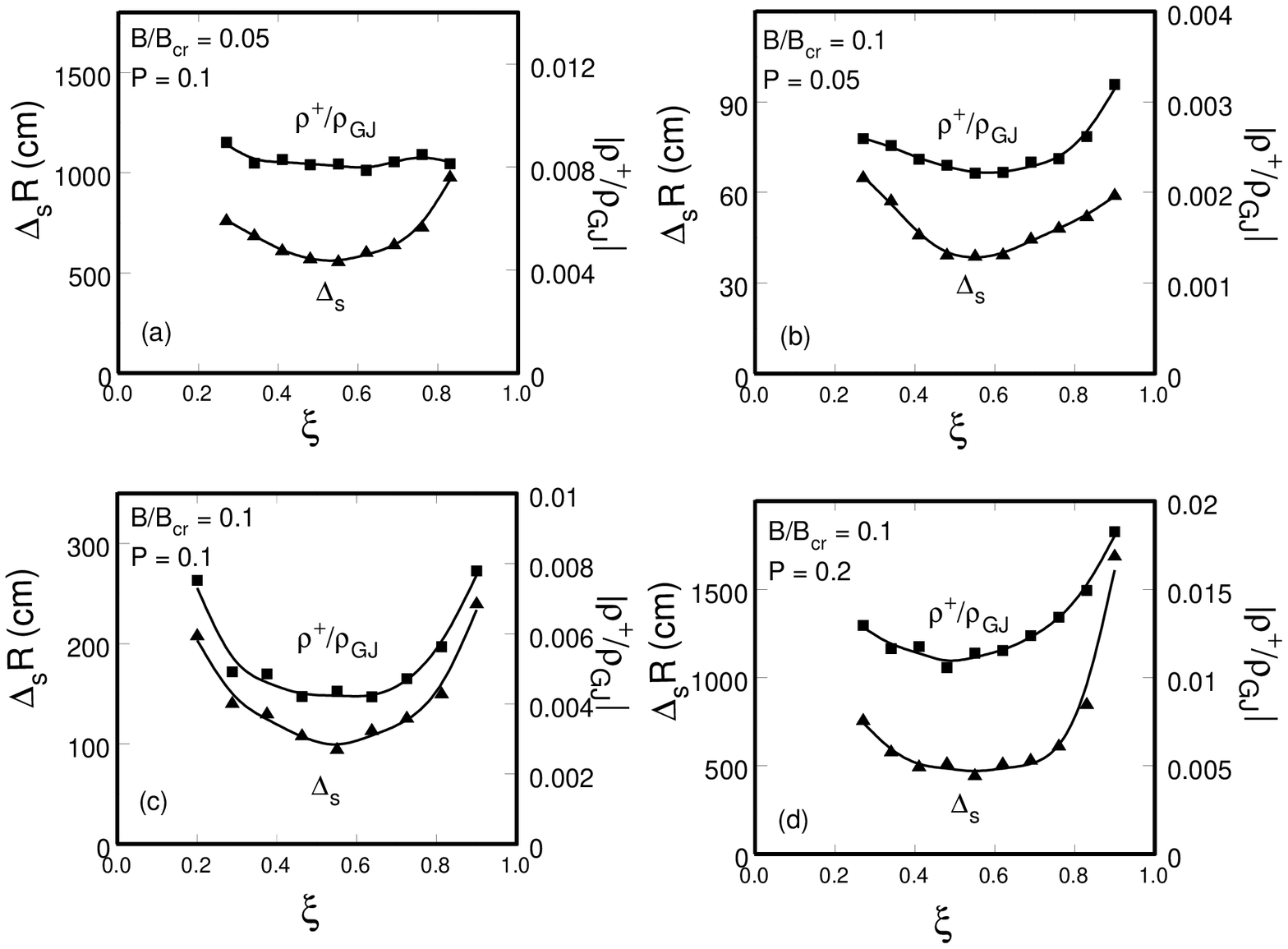}{0}{
Solutions for the returning positron density, $\rho^+/\rho_{\rm GJ}$, 
normalized to the GJ density and the screening scale height, $\Delta_s R$,
as a function of magnetic colatitude, $\xi$, which has been normalized to the
PC half-angle.
    \label{fig:} }    

\figureout{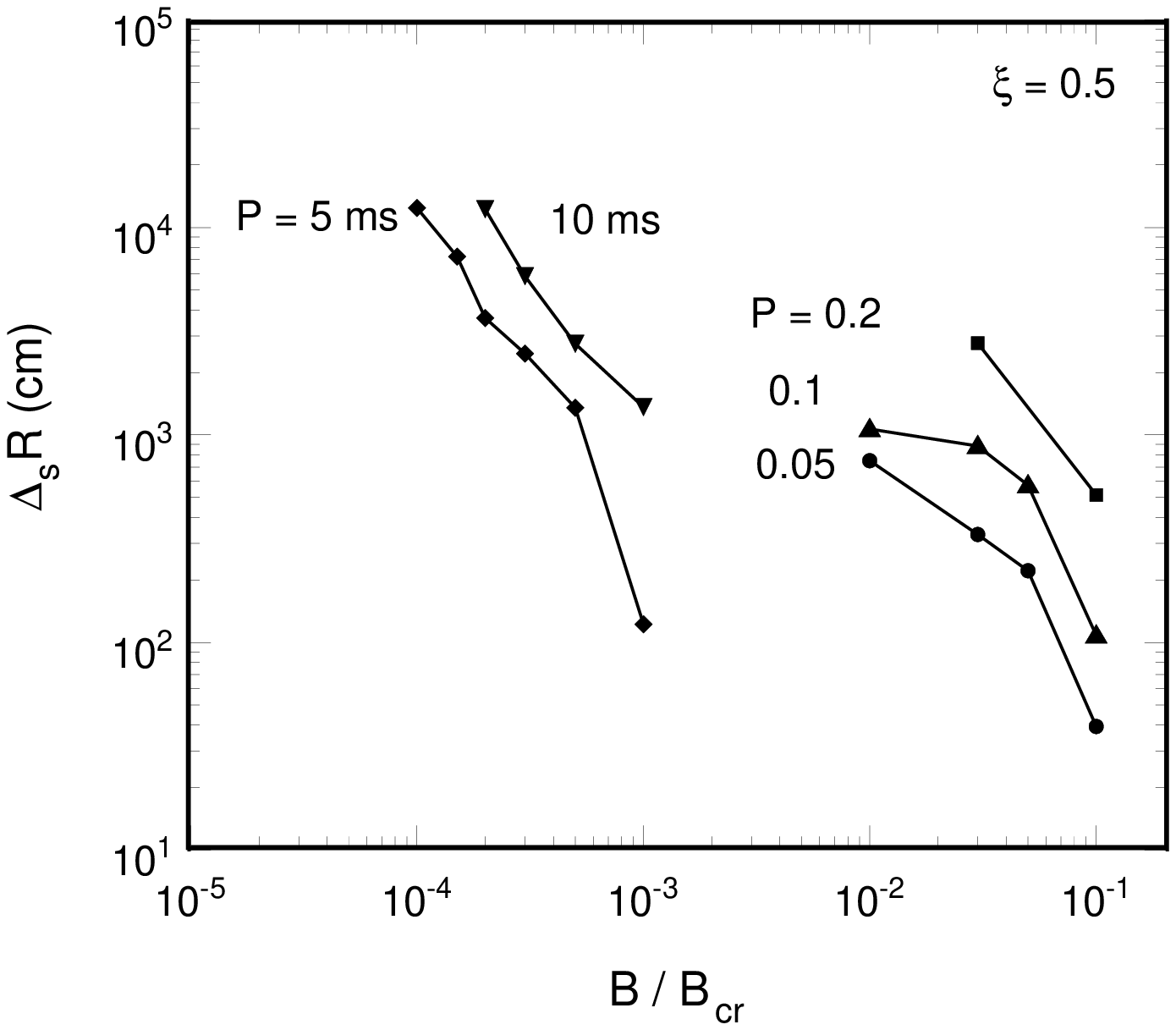}{0}{
Solutions for the screening scale height, $\Delta_s$, as a function of surface magnetic
field strength in units of the critical field, $B/B_{\rm cr}$, for different pulsar periods. 
    \label{fig:} }    

\figureout{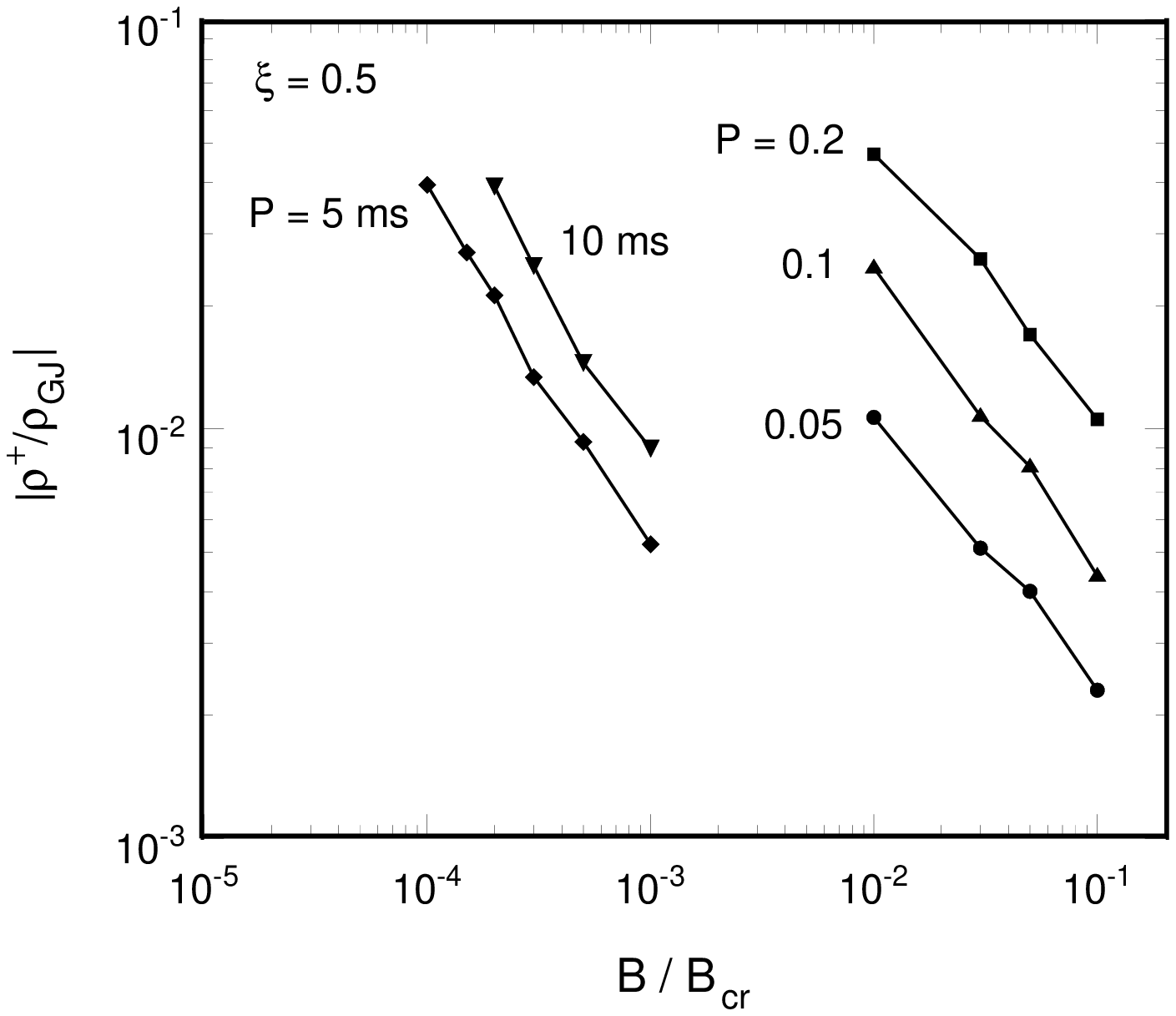}{0}{
Solutions for the returning positron density, $\rho^+/\rho_{\rm GJ}$, 
normalized to the GJ density as a function of surface magnetic field
strength in units of the critical field, $B/B_{\rm cr}$, for different pulsar periods.
    \label{fig:} }    

\figureout{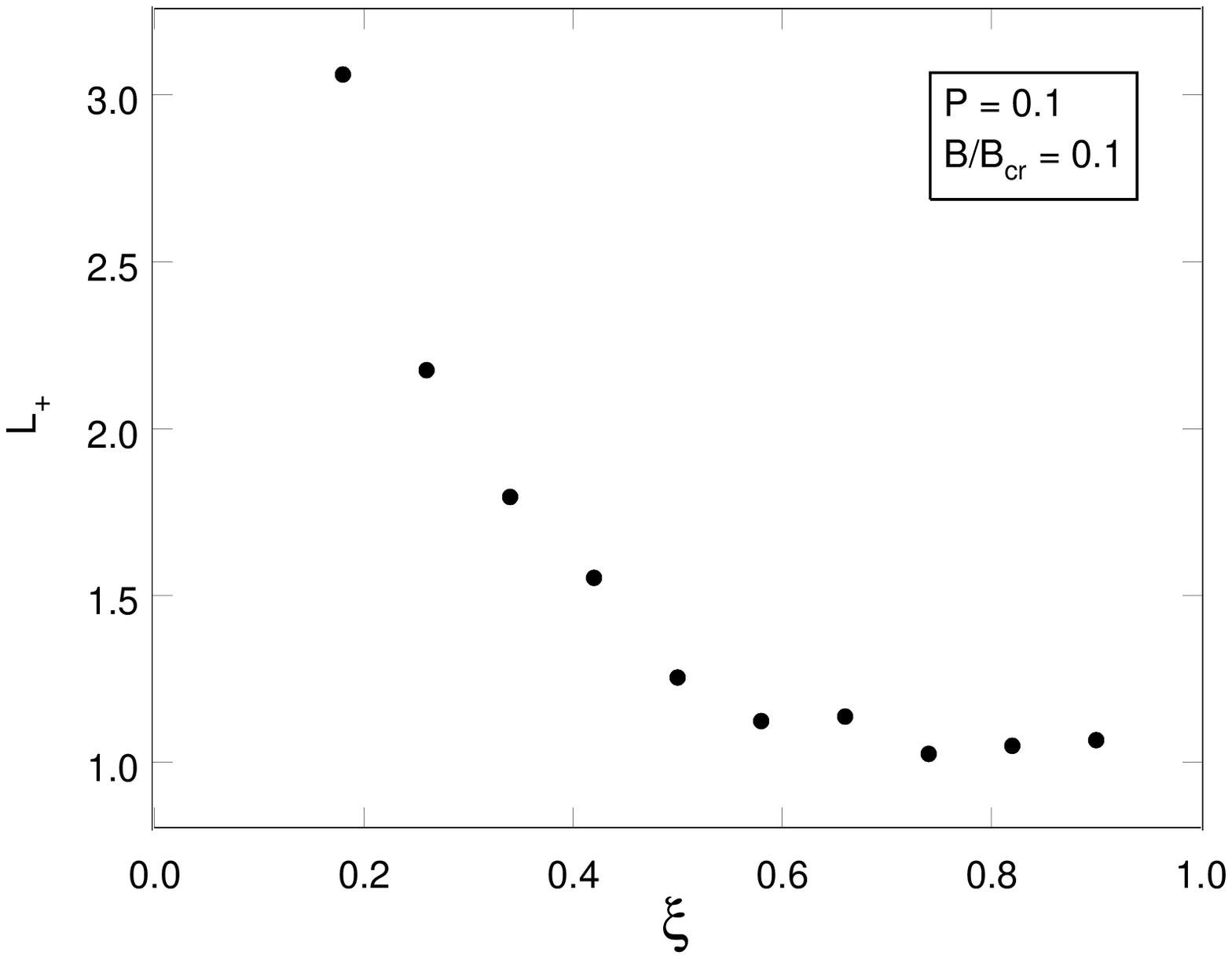}{0}{
Example of the variation of PC heating luminosity, $L_+$, as a function of  
magnetic colatitude, $\xi$, which has been normalized to the
PC half-angle.  The normalization of the vertical axis is arbitrary.
    \label{fig:} }    

\figureout{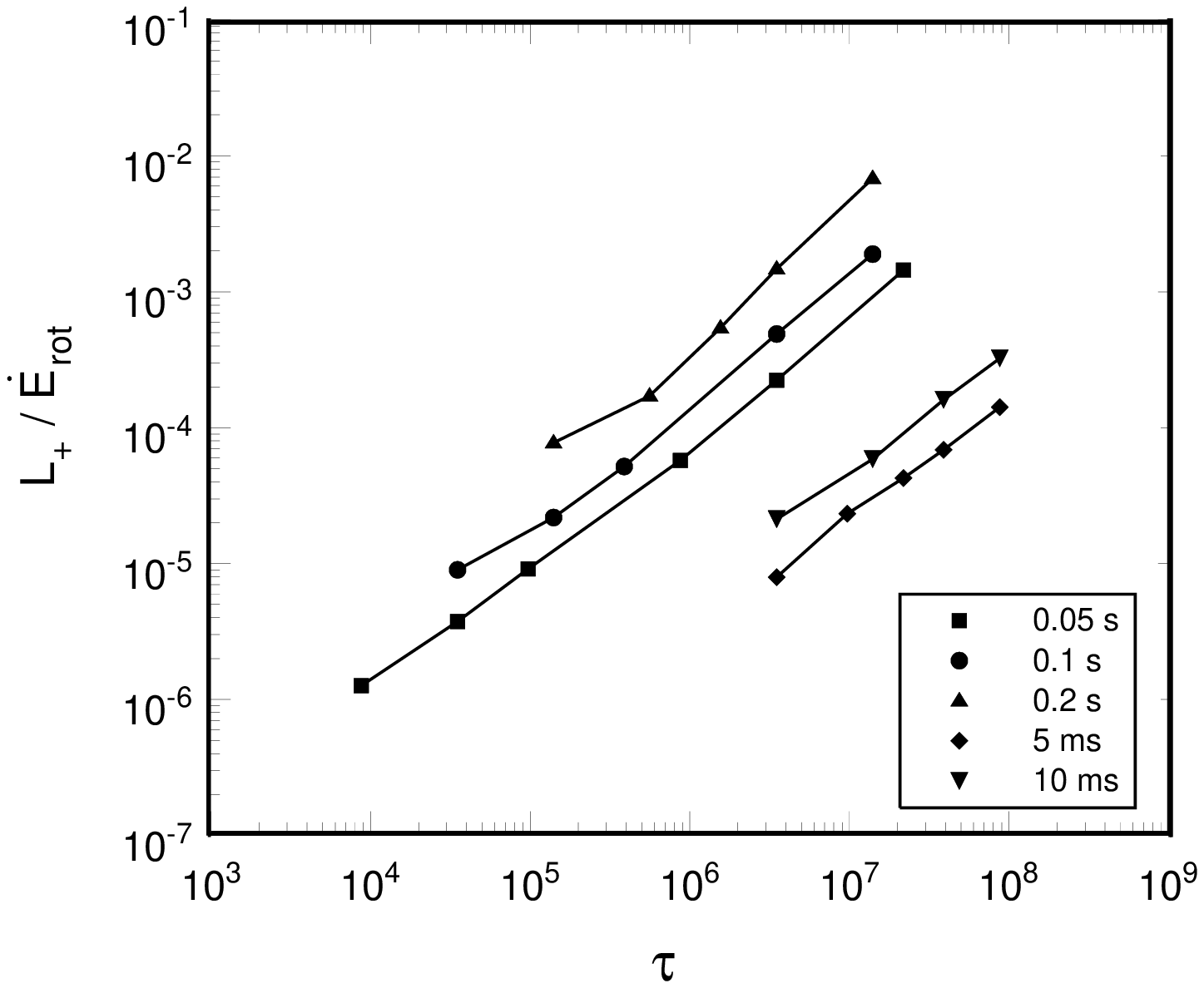}{0}{
PC heating luminosity, $L_+$, normalized to the spin-down energy loss rate,
$\dot E_{rot}$, as a function of the characteristic spin-down age, $\tau
= P/2\dot P$, for different pulsar periods, as labeled.
    \label{fig:} }    

\end{document}